\documentclass [12pt]{article}
\usepackage{latexsym}
\usepackage{amsfonts}
\usepackage{amsmath}
\usepackage{acronym}
\newcommand{\be}{\begin{equation}}
\newcommand{\ee}{\end{equation}}
\newcommand{\bea}{\begin{eqnarray}}
\newcommand{\eea}{\end{eqnarray}}
\makeatletter
\@addtoreset{equation}{section}

 \makeatother
\begin{document}
\normalsize
\title{Topological Dirac variables in Abelian $U(1)$ theory.}
\author
{{\bf L.~D.~Lantsman}\\  
Tel.  049-0381-799-07-24,\\
llantsman@freenet.de}
\maketitle
\begin {abstract}
In this study we,  remembering the experience with topological Dirac variables in the non-Abelian Yang-Mills-Higgs (YMH) model with vacuuum BPS monopole solutions, attempt to construct  similar for the Abelian $U(1)$ model. We show that QED, as one understands it commonly, is only the topologically trivial sector ($n=0$) of this Abelian $U(1)$ model.  For $n\neq 0$ one gets Dirac monopole modes. In both the cases, $n=0$ and $n\neq 0$, the theory can be quantized via the Hamiltonian reduction in terms of Dirac variables.
\end{abstract}
\noindent PACS: 12.20.m,  12.20. DS, 14.80.Hv.  \newline
Keywords: Abelian Theory, Dirac monopole.\newpage
 \tableofcontents
\newpage
\section{Introduction}
Of late, a good deal of efforts was spended to construct the constraint-shell formalism for the various Minkowskian non-Abelian models.  The essence of this method consists in the  reduction of the apropriate Hamiltonians in terms of {\it physical}, i.e.  always transverse and gauge invariant, variables, called {\it Dirac variables}.

These  Dirac variables can be got manifestly as the  solutions to the {\it Gauss law constraint}
\be \label {Gau} \partial W/\partial A_0=0\ee
(with $W$ being the action functional of the considered gauge theory and $A_0$ being the the temporal components of the apropriate gauge field 
 Dirac variables are just  such variables. There are physical fields which are solutions to the {\it Gauss law constraint} \footnote{ Generally speaking, a constraint equation in the Hamiltonian formalism  can be defined as following \cite{LP1}. Constraint equations relate initial data for
spatial components of the fields involved in a (gauge) model to initial data of their temporal components.

The Gauss law constraint has an additional specific that it is simultaneously the one of equations of motion (to solve these, it is necessary a measurement of initial
data \cite{LP1}). This is correctly for QED as well for non-Abelian theories (in particular, for QCD), i.e. for the so-called {\it particular} theories involving \cite{Gitman} the singular  Hessian matrix
$$ M_{ab}= \frac {\partial ^2 L}{\partial \dot q ^a\partial\dot q ^b} $$
(with $L$ being the Lagrangian of the studied theory,   $q ^i$ being the apropriate degrees of
freedom and $\dot q^i$ being their time derivatives).

The Hessian matrix $M$ becomes singular (i.e. ${\rm det} M=0$), sinse the identity 
$$ \partial  L/ \partial  \dot A_0 \equiv 0$$
which 'temporal' components $A_0$ of gauge fields always satisfy in particular theories (in other words, particular theories involve zero canonical momenta $\partial  L/ \partial \dot A ^0$ for  temporal components $A_0$ of gauge fields).

Thus temporal components $A ^0$ of gauge fields are, indeed, non-dynamical degrees of freedom
in particular theories, the quantization of which contradicts the Heisenberg uncertainty
principle.
}). \par 

In detail, the origin appearing Dirac variables in a particular gauge theory is following. In order to eliminate temporal components $A ^0$ of gauge fields, which are undesirable therein, Dirac \cite{Dir} and, after him, other authors of the first classical studies in quantization
of gauge fields, for instance \cite{Heisenberg,Fermi}, eliminated temporal components of gauge 
fields by gauge transformations. The typical look of such gauge transformations is \cite{David3}
\be \label{udalenie} v^T({\bf x},t)(A_0+\partial_0)(v^{T})^{-1}({\bf x},t)=0. \ee
This equation may be treated as that specifying the gauge matrices $ v^T({\bf x},t)$. This, in
turn, allows to write down the gauge transformations for spatial components of gauge fields \cite{LP1}  
\be \label{per.Diraca} {\hat A}^D_i({\bf x},t):=v^T({\bf x},t)({\hat A}_i+\partial_i)(v^{T})^{-1}({\bf x},t); \quad {\hat A}_i= g \frac {\tau ^a}{2i}A_{ai}.
\ee
It is easy to check that the functionals ${\hat A}^D_i({\bf x},t)$ specified in such a way are gauge invariant and transverse fields: 
\be \label{transvers} \partial_0  \partial_i {\hat A}^D_i({\bf x},t) =0; \quad  u ({\bf x},t)  {\hat A}^D_i({\bf x},t)   u ({\bf x},t)^{-1}= {\hat A}^D_i ({\bf x},t)  \ee
for gauge matrices $ u ({\bf x},t)$. \par 
Following Dirac \cite{Dir}, we shall refer to the functionals ${\hat A}^D_i({\bf x},t)$ as to the {\it Dirac variables}.  
The Dirac variables ${\hat A}^D_i$ may be derived by resolving the Gauss law constraint (\ref{Gau}). Solving Eq. (\ref{Gau}), one expresses  temporal components $A_0$ of gauge
fields $A$ through their  spatial components; by  that the nondynamical components $A_0$ are indeed ruled out from the apropriate Hamiltonians.  
Thus the reduction of particular gauge theories occurs over the surfaces of the apropriate Gauss law constraints. 
Only upon expressing temporal components $A_0$ of gauge
fields $A$ through their  spatial components one can perform gauge transformations (\ref{per.Diraca})  in order to  turn spatial components $\hat A_i$ of gauge
fields into gauge-invariant and transverse Dirac variables ${\hat A}^D_i$. Thus, formally, temporal components $A_0$ of these fields become zero.  
By that the Gauss law constraint (\ref{Gau}) acquires the form
\be \label{perpen} \partial_0 \left(\partial_i \hat A^D_i({\bf x},t) \right) \equiv 0.\ee

\bigskip  Such quantization method is suitable for non-Abelian as well as Abelian particular gauge theories. A good patern, how to quantize in the above wise a non-Abelian (Minkowskian) gauge model is that procedure applied to the Minkowskian theory with the $SU(2)\to U(1)$ violated gauge symmetry involving Yang-Mills (YM) and Higgs vacuum modes in the shape of BPS monopoles \cite{BPS}.

\bigskip
Here we should like to enumerate gains which the above described {\it Dirac} (or {\it fundamental}) quantization method \cite{Dir} gives for the Minkowskian gauge theory with vacuum YM and Higgs BPS monopole modes.  Among them (the author holds  it is the immediate of these gains) is the appearance of so-called {\it zero modes}, $\tilde A_0$, solutions  to the Gauss law constraint (\ref{Gau}) which can be recast to the shape of the homogeneous equation 
\be \label{koop}
(D^2)^{ab} \Phi _b ({\bf x})=0,
\ee
involving the covariant derivative $D$ of the (topologically trivial)  Higgs BPS monopole mode $\Phi _a$. 

Herewith any  zero mode $\tilde A_0$ becomes directly proportional to $\Phi _a$ with  some only time depended coefficient $\dot c(t)$:
$$\tilde A_0^a(t,{\bf x})=\dot c(t) \Phi ^a ({\bf x})\equiv Z^a.$$ 
The variable $c(t)$ is related closely to the  vacuum Chern-Simons functional:
\bea
\label{winding num.}
\nu[A_0,\Phi]&=&\frac{g^2}{16\pi^2}\int\limits_{t_{\rm in} }^{t_{\rm out} }
dt  \int d^3x F^a_{\mu\nu} \widetilde{F}^{a\mu \nu}=\frac{\alpha_s}{2\pi}  \int d^3x F^a_{i0}B_i^a(\Phi)[c(t_{\rm out}) -c(t_{\rm in})]\nonumber \\
 &&  =c(t_{\rm out}) -c(t_{\rm in})= \int\limits_{t_{\rm in} }^{t_{\rm out} } dt \dot c(t); \quad  \alpha_s\equiv g^2/4\pi.
 \eea 
 Here $F$ is the YM strenght tensor with its dual $\widetilde{F}^{a\mu \nu}=1/2 \epsilon ^{\mu \nu\lambda \rho} F_{\lambda \rho }^a$; $g$ is the YM coupling constant; $B_a^i(\Phi)=1/2 \epsilon ^{aik} F_{ik}$ ($F_{ik}=(\Phi^a/\vert\Phi\vert) \vert F_{ik a}$) is the "magnetic" field generated vacuum BPS monopole modes $\Phi$ via the {\it Bogomol'nyi equation} \cite{BPS,Gold,Al.S.}
 \be
\label{Bog}
 {\bf B} =\pm D \Phi. 
\ee

Zero modes $\tilde A_0^a(t,{\bf x})$ generate vacuum "electric" strength
\be
\label{el.m}
F^a_{i0}={\dot c}(t)D ^{ac}_i(\Phi )\Phi_{c}({\bf x})
\ee 
referred to as the so-called  "electric monopole" (for instance, in  \cite{LP1,LP2}).

In turn, "electric monopole" modes $F^a_{i0}\equiv E^a$ enter, in a natural wise,  the action functional 
\be \label{rot} W_N=\int d^4x \frac {1}{2}(F_{0i}^c)^2 =\int dt\frac {{\dot c}^2 I}{2},\ee
involving the  ``rotary momentum''
\be \label{I} I=\int \sb {V} d^3x (D_i^{ac}(\Phi_k)\Phi_{c})^2.    \ee
Thus, as it was argued in \cite{LP1,LP2, David2}, this action functional (\ref{rot}) describes correctly  {\it collective solid rotations} inside the YM-Higgs (further, YMH) vacuum involving BPS monopole modes and quantized by Dirac in the above wise. The remarkable property of the  action functional (\ref{rot}) is also the purely real, i.e. physical  spectrum 
\be \label{pin} P_c ={\dot c} I= 2\pi k +\theta; \quad \theta  \in [-\pi,\pi]; \quad k\in {\bf Z}     \ee
of the topological momentum $P_c$.

In this  purely real spectrum is the principal distinction of the {\it Minkowskian} YMH model (with vacuum BPS monopole modes) considered in the  Dirac quantization scheme \cite{Dir} from the {\it Euclidian} YM theory involving instantons \cite{Bel}.  In the latter theory, as it was discussed in the papers \cite{Pervush1,Galperin,rem3}, the $\theta$-angle, playing the role of a quasi-momontum therein, takes indeed a complex value $\theta=\theta_1+i\theta_2$. This implies the indefinite norm for quantum objects corresponding instantons \cite{Bel}, and thus this model encounters lot of problems.

\bigskip The manifest rotary effect (\ref{rot}) inherent in the Minkowskian YMH model with vacuum BPS monopoles quantized by Dirac distinguishes that model, to a marked degree, from the well-known 't Hooft-Polyakov model (also associated with the Minkowski space) involving of the same name vacuum monopoles \cite{H-mon,Polyakov}. The latter model does not contain rotary (vacuum) modes, but only stationary solutions,   monopoles \cite{H-mon,Polyakov}.

The said, indeed, is a specific trace of the {\it heuristical} quantization scheme by Faddev and Popov (further, FP) \cite{FP1} involving the relativistic (Poincare) invariant S-matrix and only mass-shell quantum fields. This quantization scheme \cite{FP1} is not attached to a definite reference frame. On the contrary, the Dirac quantization scheme \cite{Dir} is always associated with a definite reference frame, for instance with the rest reference frame. Just in such rest reference frame the rotary effect (\ref{rot}) can be observed in the  Minkowskian YMH model with vacuum BPS monopoles quantized by Dirac, while the S-matrix aproach \cite{FP1} to the 't Hooft-Polyakov model \cite{H-mon,Polyakov} gives only statical (stationary) solutions.

\medskip The next important distinction between the 't Hooft-Polyakov model \cite{H-mon,Polyakov} and the Minkowskian YMH model with vacuum BPS monopoles quantized by Dirac lies in thermodynamics.  As it is well known, the second order phase transition occurs in the former theory and is reduced to the spontaneous and {\it instantaneous} breakdown of the initial $SU(2)$ gauge symmetry group up to its $U(1)$ subgroup. In the Minkowskian YMH model with vacuum BPS monopoles quantized by Dirac the {\it first order phase transition} takes place \footnote{Side by side with the above second order phase transition involving $SU(2)\to U(1)$.} coming to coexisting (at the absolute zero temperature $T=0$) two thermodynamic phases inside the BPS monopole vacuum. 

The first of these two thermodynamic phases is the phase of collective vacuum rotations described by the action functional  (\ref {rot}).  The second one is the thermodynamic phase of superfluid potential motions set by the Bogomol'nyi equation (\ref {Bog}) and the {\it Gribov ambiguity  equation} 
\footnote{The origin of the Gribov ambiguity  equation (Gribov   equation) (\ref{Gribov.eq}) is following. In the Minkowskian YMH model with vacuum BPS monopoles quantized by Dirac the general expression (\ref{per.Diraca}) for the Dirac variables  acquire the concrete look \cite{LP1,LP2,David2}
\bea \nonumber
\hat A_k^D = v^{(n)}({\bf x})T \exp \left\{\int  \limits_{t_0}^t d {\bar t}\hat A _0(\bar t,  {\bf x})\right\}\left({\hat A}_k^{(0)}+\partial_k\right ) \left[v^{(n)}({\bf x}) T \exp \left\{\int  \limits_{t_0}^t d {\bar t} \hat A _0(\bar t,{\bf x})\right\}\right]^{-1}, \eea 
with  the symbol $T$  standing for   time ordering  the matrices under the exponent sign. \par
Thus in the initial time instant $t_0$, the topological degeneration of initial (YM) data
comes thus to  "large" stationary matrices $v^{(n)}({\bf x})$ ($n\neq 0$) [in the terminology \cite {Jack}] depending on  topological numbers $n\neq 0$ and called the factors of the
Gribov topological degeneration or simply the \it Gribov multipliers\rm.  \par
One attempts \cite {LP1,LP2, Pervush2} to find Gribov multipliers $v^{(n)}({\bf x})$, belonging to the $U(1)\subset SU(2)$ embedding in the Minkowskian Higgs model, as
$$ \exp [n \hat\Phi _0({\bf x})],$$
implicating the {\it Gribov phase} $\hat\Phi _0({\bf x})$, taking the shape \cite {David3} of a scalar constructed by contracting the Pauli matrices $\tau^a$ and Higgs vacuum BPS monopole modes: 
$$ 
{\hat \Phi}_0(r)= -i\pi \frac {\tau ^a x_a}{r}f_{01}^{BPS}(r), \quad 
f_{01}^{BPS}(r)=[\frac{1}{\tanh (r/\epsilon)}-\frac{\epsilon}{r} ].$$ 
In the initial time instant $t_0$ the topological Dirac variables $\hat A_k^D$ acquire the look
$$ {\hat A}^{(n)}_k= v^{(n)}({\bf x}) ({\hat A}_k^{ (0)}+
\partial _k)v^{(n)}({\bf x})^{-1},\quad v^{(n)}({\bf x})=
\exp [n\Phi _0({\bf x})].$$
The important property of these topological  is their "transverse" character: namely, that \cite {Pervush2}
$$D_i^{ab} (\Phi _k^{(n)}){\hat A}^{i(n)}_b =0 $$
with the covariant derivative $D$ depending on the (topologically degenerated) vacuum YM BPS monopole modes $\Phi _k^{(n)}$ (indeed, this is controlled by the Bogomol'nyi equation (\ref{Bog})).

The above transverse gauge for the topological Dirac variables $\hat A_k^D$ is not specified in a unique wise in each topological sector §n§ of the considered YMH model. This fenomenon (correct for any transverse gauge in any non-Abelian gauge theory \cite{Al.S.}) is referred to as the {\it Gribov ambiguity}. The general analysis of this effect in the terminology of the trivial principal fibre bundle (involving the gauge group $SU(2)$) was given in the \S T26 in the monograph \cite{Al.S.}, which we recommend to our readers.

\medskip In our concrete Minkowskian YMH model with vacuum BPS monopoles quantized by Dirac the Gribov ambiguity fenomenon for   (topologically degenerated) Dirac variables $\hat A_k^{D(n)}$ just comes to the  Gribov ambiguity  equation  (\ref{Gribov.eq}). To ground this, it is necessary to write down explicitly the YM "magnetic" field
$$B_i^a= \epsilon_{ijk} (\partial^j A^{ak} +\frac {g}{2}\epsilon ^{abc}A_{b}^j A_{c}^k). $$
Then it is easy to see that the values $D_i A^{ia}$ (in particular, $D_i A^{i D}$ if topological Dirac variables $A^D$ are in question) have the same dimension  that a "magnetic" YM field $ B_i^a$. Then it is easy to see that the Gribov ambiguity equation (\ref{Gribov.eq}) is the consequence of the Bogomol'nyi equation (\ref{Bog}), implicating (topologically trivial) Higgs vacuum BPS monopole modes $\Phi_{(0)}$. Precisely, the connection between the Bogomol'nyi and Gribov ambiguity equations is set through the  Bianchi identity
$$\epsilon ^{ijk}\nabla _i F_{jk}^b =0,$$ 
that is equivalent to Eq.
$$ D B=0$$
in terms of the (vacuum) "magnetic"  field $\bf B$.
}  
\be
\label{Gribov.eq} [D^2 _i(\Phi _k]^{ab}\Phi_{b} =0.
\ee
\medskip This coexisting at the temperature $T=0$ of two thermodynamic phases can continue infinitely long time until $T=0$. Such first order phase transition was discussed in the recent paper \cite{disc} (see Conclusion therein), where it was referred to as the {\it frozen supercooling situation}. Herewith collective solid rotations inside the BPS monopole vacuum proceed without "friction forces". In this is the essence of the {\it Josephson effect} \cite{ rem3,Josephson,Pervush3}: at $T=0$, any "quantum train" cannot stop, moving permanently along closed trajectories.

\medskip The first order phase transition occurring in the Minkowskian YMH model with BPS monopole vacuum quantized by Dirac finds its reflection in the vacuum (Bose condensation) Hamiltonian $H_{\rm cond}$  \cite{LP1,LP2}
\be
\label{Hamilton}
H_{\rm cond}= \frac {2\pi}{g^2\epsilon}[ P_c^2 (\frac {g^2}{8\pi^2})^2+1],
\ee
written down over the YM Gauss law constraint (\ref{Gau}) and containing the "electric" and "magnetic" contributions, given via Eqs. (\ref{rot}) and 
\be 
\label{magn.e1}
\frac{1}{2}\int \limits_{\epsilon}^{\infty } d^3x [B_i ^a(\Phi_k)]^2 \equiv \frac{1}{2}V <B^2> =\frac 1{2\alpha_s}\int\limits_{\epsilon}^{\infty}\frac {dr}{r^2}\sim \frac 1 2  \frac 1
{\alpha_s\epsilon}= 2\pi \frac{gm}{g^2\sqrt{\lambda}}=\frac {2\pi} {g^2\epsilon}, 
\ee
respectively. The latter,  "magnetic", contribution is associated with the Bogomol'nyi equation (\ref {Bog}). 

The remarkable property of the vacuum (Bose condensation) Hamiltonian $H_{\rm cond}$ is its {\it Poincare invariance} (and thus also the CP-invariance) as that squared by the topological momentum $P_{c}$. This solves the CP-problem (taking place in the {\it Euclidian} insanton model \cite{Bel} involving the Poincare covariant $\theta$-item in its Lagrangian \footnote{We recomend our readers the paper \cite{rem3} where the instanton model \cite{Bel} was stated enough briefly but informatively.}) in the {\it Minkowskian} YMH theory with vacuum BPS monopoles quantized by Dirac.

\bigskip To explain the above (frozen) first order phase transition taking place in the Minkowskian YMH model with BPS monopole vacuum quantized by Dirac, the special assumption about the $SU(2)\to U(1)$ gauge group space inherent in that  model was made in the recent papers \cite{disc,fund}. 

The essence of this assumption is in the so-called "discrete" factorization
\be
 \label{fact2}
SU(2)\equiv G\simeq G_0\otimes {\bf Z}; \quad     U(1)\equiv H \simeq U_0 \otimes {\bf Z}
 \ee
 of the initial, $SU(2)$, and residual, $U(1)$, gauge symmetries groups.
 
 In this case it can be shown that the apropriate YM degeneration space (vacuum manifold) $R_{\rm YM}\equiv SU(2)/ U(1)$ acquires the look
  \be
 \label{RYM}
 R_{\rm YM}= {\bf Z}\otimes G_0/U_0. \ee
 $R_{\rm YM}$ is a manifestly multiconnected  (discrete) space: $\pi_0 (R_{\rm YM})={Z}$. This means that different topological sectors of  $R_{\rm YM}$ are separated by {\it domain walls} \footnote{As it is well known (see, e.g. Ref. \cite{Ph.tr}), the width of a domain (or \it Bloch\rm, in the terminology  \cite{Ph.tr}) wall is roughly proportional to the inverse of the lowest mass of all the physical particles in the (gauge) model considered. \par 
 In Minkowskian Higgs models (without quarks) the typical such scale is the (effective) Higgs mass $m/\sqrt\lambda$ (with $m$ being the Higgs mass and $\lambda$ being its selfinteraction constant). In particular, in the Minkowskian Higgs model with vacuum BPS monopoles quantized by Dirac (we discuss now) $m/\sqrt\lambda$ is the only mass scale different from zero in the Bogomol'nyi-Prasad-Sommerfeld (BPS) limit \cite{BPS}
 
 $$m \to 0  \quad \lambda \to 0.$$ If quarks are incorporated nevertheless in this model, one thinks that any ``bare'' flavour mass $m_0$ is by far less than the ``effective'' Higgs mass $m/\sqrt\lambda$: 
$$ m_0\ll m/\sqrt\lambda. $$
The typical value  of the length dimension inversely proportional to $m/\sqrt\lambda$ is the (typical) size $\epsilon$ of BPS monopoles.

It can be given as  \cite{LP1,  LP2,David2}
$$\frac{1}{\epsilon}=\frac{gm}{\sqrt{\lambda}}\sim \frac{g^2<B^2>V}{4\pi}, $$
with $V$ being the volume ocuppied by the YMH vacuum configuration. 

The said allows to assert that $\epsilon$ disappears in the infinite  spatial volume limit $V\to\infty$, while it is maximal at the origin of coordinates (herewith it can be set $ \epsilon (0)\to\infty$).
This means, due to the above reasoning \cite{Ph.tr}, that walls between topological domains inside $R_{\rm YM}$ become infinitely wide, $O(\epsilon (0))\to\infty$, at the origin of coordinates. 

The fact $\epsilon(\infty)\to 0$ is  also meaningful.  This implies  actual merging of topological domains inside the vacuum manifold $R_{\rm YM}$, (\ref{RYM}), at the spatial infinity. The said allows \cite{disc} to interpret the discrete space $R_{\rm YM}$ as the Riemann surface for the function ${\rm lim }~_{n\to \infty} (1/(\sqrt z)^n$ of the complex variable $z$ (with natural setting $V={\rm Re}~ z$). $z\to \infty$ serves as the branching point for this  Riemann surface  on which the above limit turns into zero, while   $z\to 0$, the pole point for $1/(\sqrt z)^n$ at any $n$, can be  considered as another branching point.
 }.
 
   And moreover, the YM degeneration space $R_{\rm YM}$, (\ref{RYM}), contains {\it thread} and {\it point} topological defects.
   
   Thread topological defects (vortices) inside $R_{\rm YM}$ are induced by the manifest isomorphism \cite{Al.S.}
   \be \label{iso} \pi_1 (R_{\rm YM})= \pi_0 (H)\neq 0.\ee 
   It is easy to see that these are just responsible for all the (vacuum) rotary effects inherent in the  Minkowskian Higgs model with vacuum BPS monopoles quantized by Dirac (in particular for the action functional (\ref{rot})) \footnote{Indeed, vortices arising inside the vacuum manifold $R_{\rm YM}$ are concentrated, in the main, in the spatial region along the axis $z$ of the chosen rest reference frame, infinitely close to this axis. In other words, it is just the above discussed limit $V\to 0$ ($\epsilon \to\infty$).
   
   In this spatial region vacuum  vortices can be represented \cite{Al.S.,disc} by the YM fields
   $$ A  _\theta(\rho, \theta, z) \equiv A_\mu \partial x^\mu/ \partial \theta=\exp(iM\theta) A  _\theta (\rho) \exp(-iM\theta),$$
  with $M$ being the generator of the group $G_1$ of rigid rotations compensating changes in the vacuum (Higgs-YM) ``thread'' configuration $(\Phi^a,A_\mu^a)$ at rotations around the  axis $z$ of the chosen (rest) reference frame. 
  
  Herewith
  $$  A  _\theta (\rho) = M+ \beta (\rho),
  $$
where the function $\beta (\rho)$ aproaches zero as $\rho \to \infty$.  \par  
 The elements of $G_1$ can be set as \cite{Al.S.}
 $$ g_\theta =\exp(iM\theta). $$ 
 YM fields $A_\theta $ are manifestly invariant with respect to shifts along the axis $z$.
 
 In turn, the Higgs rotary (vortex) modes $\Phi^a$ can be represented as \cite{Al.S.}
 $$  \Phi^{(n)} (\rho, \theta, z)= \exp (M\theta)~ \phi  (\rho) \quad (n\in {\bf Z}), \quad \nabla_\mu \phi(\rho) \leq {\rm const}~\rho ^{-1-\delta}; \quad  \delta>0.$$
 These solutions are singular at $\rho    \to 0$ but disappear as $\rho    \to \infty$. This property of the Higgses $\Phi^{(n)} (\rho, \theta, z)$ allows to join them (in a smooth wise) with vacuum Higgs BPS monopoles belonging to the same  topology $n$ and disappearing \cite{BPS,Al.S.,David3} at the origin of coordinates. Herewith, speaking "in a smooth wise", we imply that the covariant derivative $D\Phi$ of any vacuum Higgs field $\Phi_a^{(n)}$ merges with the covariant derivative of such a vacuum Higgs BPS monopole solution. \par
 Following \cite{Al.S.}, the vacuum $z$-invariant, i.e. {\it axially symmetric}, (Higgs-YM) configuration $(\Phi^a,A_\mu^a)$, possessing, as it can be demonstrated \cite{Al.S.}, a finite linear energy density and obeying the apropriate equations of motions) can be treated as a rectilinear thread solution (called also {\it the rectilinear thread vortex} or {\it the rectilinear thread}). \par 
 Obvious locating of (topologically nontrivial) threads $A_ \theta$ at the origin of coordinates (actually, in the spatial region $\rho \to 0$; the same is correctly also for Higgs thread modes $\Phi^a $) permits the natural  geometrical interpretation of (topologically nontrivial) threads as infinitely narrow tubes around the axis $z$ over which the family of vortex solutions   $(\Phi^a,A_\mu^a)$ is specified actually (disappearing rapidly outside this spatial region). \par
 
All  the said provides that the action functional (\ref{rot}), involving $D_i^{ac}(\Phi_k)\Phi_{c}$, is described correctly by  Higgs vacuum BPS monopole modes \cite{BPS} as well as by "rotary" Higgs modes $\Phi^{(n)} (\rho, \theta, z)$ \cite{Al.S.}. And moreover,  the topological momentum $P_c (n)$, (\ref{pin}), of the Minkowskian  BPS monopole vacuum, running over the set $\bf Z$ of integers, takes the unique value $P_c (k)=\theta +2\pi k$ in each topological sector $k$ of the Minkovskian degeneration space $R_{\rm YM}$, corresponding to the family of vortex solutions with the topological number $k$.
 }.
 
 \bigskip Point topological defects inside the YM degeneration space $R_{\rm YM}$ are generated by the isomorphism 
  \be \label {hedg} \pi_2 (R_{\rm YM})= \pi_1 (H)={\bf Z},\ee 
  grounded in Ref. \cite{disc}. These topological defects come inside  $R_{\rm YM}$ to {\it point hedgehog topological defects} in the shape of vacuum Higgs and YM BPS monopoles \cite{BPS}. It is obvious that these topological defects are responsible for all the superfluidity effects controlled by the Bogomol'nyi, (\ref{Bog}), and Gribov ambiguity, (\ref{Gribov.eq}), equations  in the Minkowskian YMH model with BPS monopole vacuum quantized by Dirac. 
  
 \bigskip  To finish our survey about the  Minkowskian YMH model with BPS monopole vacuum quantized by Dirac, we should like to point of the peculiarities of QCD based on such model.
 
 1. The above discussed \cite{disc} "discrete vacuum geometry" (\ref{RYM}) of the vacuum manifold $R_{\rm YM}$ (reduced to the ${\rm lim }~_{n\to \infty} (1/(\sqrt z)^n$ Riemann surface), with  merging topological domains at the spatial infinity, promotes the specific effect, the so-called {\it topological confinement}, coming \cite{Pervush2} to decoupling from the real momentum spectrum $P_c$, (\ref{pin}), of the free rotator (\ref{rot}) the series of values $p\neq 2\pi k+\theta$ ($k\in {\bf Z}$) \footnote{This becomes transparent if one considers the wave function
 $$
 \Psi (c)=e^{ipc},
 $$
 corresponding to the free rotator (\ref{rot}). If one averages this function over all the values $n\in \bf Z$ of the topological degeneration
 with the $\theta$-angle measure $\exp (i\theta n)$, we get \cite{Pervush2} 
 $$ \Psi (c)_{\rm observable}=\lim\limits_{L \to \infty}\frac{1}{2L}
 \sum\limits_{n=-L }^{n=+L }
 e^{i\theta n}\Psi (c+n)=\exp\{i (2\pi k+\theta)c\}. $$
 It is so since an observer does not know where is the rotator (\ref{rot}).  It can be at points
 $N_{in},N_{in}\pm 1,N_{in}\pm 2,N_{in}\pm 3,\dots$
 
 At deriving $\Psi (c)_{\rm observable}$ the relation \cite{Pervush3} 
 $$ \frac 1 L \sum \limits_ {n=-L/2}^ {n=L/2} =1$$
 was utilizerd.
 
 Thus we see that if $p\not = 2\pi k + \theta$, $\Psi (c)_{\rm observable}=0$. Just this fenomenon was referred to as the {\it complete destructive interference} in Refs. \cite{LP1,LP2,Pervush2}. }. The physical sense of the topological confinement comes to surviving, in the vacuum Hamiltonian (\ref{Hamilton}),  the set $\bf Z$ of integers, entering this Hamiltonian via the topological momentum $P_c$, (\ref{pin}) inspite the manifest gauge invariance of this vacuum Hamiltonian  due to the absorption of the Gribov topological multipliers $ v^{(n)}({\bf x})$ therein.
 
 \bigskip 2. The topological confinement, in the spirit of the complete destructive interference \cite{LP1,LP2,Pervush2} of the  topologically nontrivial Gribov  multipliers $ v^{(n)}({\bf x})$ ($n\neq 0$), implies the quark confinement as it is understood customary: one cannot observe colored quarks, i.e. those "dressed" in Gribov topological multipliers $ v^{(n)}({\bf x})$ ($n\neq 0$):
 $$q^{\rm I} v^{(n)}({\bf x})$$
 ($q^{\rm I}$ are topologically trivial  quarks, which are gauge invariant, that means they are  colorless).
 
 At the QCD Hamiltonian level this implies its gauge invariance \cite{Pervush2}:
 \be \label{invham}H[A^{(n)},q^{(n)}]=H[A^{(0)},q^{\rm I}]\ee
 ($A^{(n)}$ are gluonic fields contained the Gribov topologically nontrivial multipliers $ v^{(n)}({\bf x})$, $n\neq 0$).
 
 And moreover, as it was shown in  Ref. \cite{Azimov}, in the lowest order of the perturbation theory, averaging (quark) Green functions over all  topologically nontrivial field configurations (including vacuum monopole ones) prove to be \cite{Azimov,Nguen2} 
 \be \label{oneqw}
G({\bf x},{\bf y})= \frac {\delta}{\delta s^*(x)} \frac {\delta}{\delta \bar
s^*(y)} Z_{\rm conf} (s^*, \bar s^*,J^*)\vert_{ s^*= \bar s^*=0} =G_0(x-y) f({\bf x},{\bf y}),
\ee
with $ G_0(x-y)$ being  the (one-particle) quark propagator in the perturbation theory and
\be \label{interf}
f({\bf x},{\bf y})= \lim_{\vert {\bf x}\vert \to \infty, ~ \vert {\bf y}\vert \to \infty} \lim_{L \to \infty} (1/L) \sum \limits _{n=-L/2}^{n=L/2} v^{(n)} ({\bf x}) v^{(n)} ({\bf -y}). \ee
Further, $s^*$, $\bar s^*$,  $J^*$ are the sources of the quark ($q$), antiquark ($\bar q$) and gluonic ($A$) fields, respectively, while $ Z_{\rm conf} (s^*, \bar s^*,J^*)\vert_{ s^*= \bar s^*=0}$ is the generating functional given in the transverse gauge
 \footnote{Fixing the gauge (\ref{tg}) implies the Faddeev-Popov (FP) integral in the shape \cite{Azimov}
$$
Z_{R,T} (s^*, \bar s^*,J^*)= \int D A_i^* Dq^* D \bar q^* ~{\rm det}~ \hat \Delta ~\delta (\int \limits_ {t_0}^t d\bar t D_i (A)\partial_0 A^i)
$$
\be \nonumber \label{Zr}
\times
\exp \{ i \int \limits_ {-T/2}^ {T/2} dt \int \limits_ {\vert {\bf x}\vert \leq R} d^3x [{\cal L} ^I (A^*, q^*) +  \bar s^*q^* +\bar q^* s^* +J^{*a}_i A^{i*}_a]\}, 
\ee 
involving the FP operator \cite{Baal}
$$\hat \Delta = -(\partial_i D^i (A)) = -( \partial_i^2 + \partial_i~ {\rm ad} (A^i))$$
with
$$ {\rm ad} (A) X \equiv [A,X]$$ 
for an element $X$ of the apropriate Lee algebra. 

The very important feature of the FP integral $Z_{R,T} (s^*, \bar s^*,J^*)$ is its expressing in terms of Dirac variables ($A_i^*$, $q^*$, $\bar q^*$), i.e. its actual dependence on Gribov topological multipliers $ v^{(n)}({\bf x})$.

With loss of generality, one can set $T\to \infty$. The FP integral $Z_{R,T} (s^*, \bar s^*,J^*)$ includes the Lagrangian density ${\cal L}^I$ corresponding to the constraint-shall action of the Minkowskian non-Abelian theory (Minkowskian QCD) taking on the surface of the Gauss law constraint (\ref {Gau}); also  $R$ is a large real number, and one can assume  that $R\to \infty$.

\medskip Then the generating functional $Z_{\rm conf} (s^*, \bar s^*,J^*)$ in Eq. (\ref {oneqw}) may be derived from the FP integral $Z_{R,T} (s^*, \bar s^*,J^*)$ by its averaging over the Gribov topological degeneration of initial data, i.e. over the set $\bf Z$ of integers
\be \label{Zcon}\nonumber
Z_{\rm conf} (s^*, \bar s^*,J^*)= \lim _{\vert {\bf x}\vert\to \infty,~T\to \infty} \lim _{L\to \infty} \frac{1}{L} \sum \limits _{n=-L/2}^{n=L/2}  Z_{R,T}^I (s^*_{n,\phi_i}, \bar s^*_{n,\phi_i},J^*_{n,\phi_i}), \ee 
}
\be \label{tg} D_i (A) \partial_0A^i =0,  \ee .

\medskip Indeed, it turns out that $ f({\bf x},{\bf y})=1$ in (\ref{oneqw}) \cite{Azimov}.  It is so due to the spatial asymptotic \cite{Pervush3,Azimov,Arsen}
\be \label{bondari} v^{(n)}({\bf x})\to \pm 1, \quad {\rm as} ~~\vert {\bf x}\vert \to \infty\ee
of the Gribov topological multipliers $ v^{(n)}({\bf x})$ (at deriving the relation $ f({\bf x},{\bf y})=1$  the same arguments as at deriving $\Psi (c)_{\rm observable}$ above were utilized) \footnote{An of no small importance circumstance promoting the spatial assymptotic (\ref{bondari}) for the Gribov  multipliers $ v^{(n)}({\bf x})$, equal for different  topologies $n$ inside the vacuum manifold $R_{\rm YM}$ is decreasing (in effect down to zero), in this limit, the widths of domain walls between  different topological sectors of this  manifold.}.

Thus we see that in the lowest order of the perturbation theory any Green function $G({\bf x},{\bf y})$ becomes topologically trivial. And this means simultaneously the topological confinement and the quark in its generally acepted sense.

\bigskip 3. New interesting properties acquire  fermionic (quark) degrees of freedom $q^*$, $\bar q^*$  in Minkowskian constraint-shell QCD involving the spontaneous breakdown of the initial $SU(3)_{\rm col}$ gauge symmetry in the
\be \label{break} SU(3)_{\rm col} \to SU(2)_{\rm col} \to U(1) \ee
way.

The only specific of  Minkowskian constraint-shell QCD (in comparison with the constraint-shell Minkowskian (YM-Higgs) theory) is the presence  of three Gell-Mann matrices $\lambda^a$, generators of $ SU(2)_{\rm col}$ (just these matrices would enter G-invariant quark currents $ j_\mu^{Ia}$ in of  Minkowskian constraint-shell QCD). 
In the constraint-shell Minkowskian (YM-Higgs) theory, involving the initial $ SU(2)$ gauge symmetry, the Pauli matrices $\tau^a$ ($a=1,2,3$) would replace the Gell-Mann $\lambda^a$ ones.

The very interesting situation, implying lot of important consequences, takes place to be in Minkowskian constraint-shell QCD involving the spontaneous breakdown (\ref{break}) of the initial $SU(3)_{\rm col}$ gauge symmetry  when the antisymmetric 
Gell-Mann matrices 
\be \label{choice} \lambda_2,\lambda_5, \lambda _7 \ee  
are chosen to be the generators of the
$SU(2)_{\rm col}$ subgroup in (\ref{break}), as it was done in Refs. \cite{David3,Pervush2,David1}. 

As it was demonstrated in  \cite{David3}, the "magnetic" vacuum field $ B^{i a}(\Phi_i)$ corresponding to Wu-Yang monopoles $\Phi_i $ \cite{Wu} acquires the form
\be \label{b11}
b_i^a=\frac 1 g \epsilon_{iak}\frac{n_k(\Omega)}{ r}; \quad
n_k(\Omega)=\frac{x^l\Omega_{lk}}{r}, \quad n_k(\Omega) n^k(\Omega)=1;
\ee 
in terms of  the antisymmetric 
Gell-Mann matrices $\lambda_2, \lambda_5, \lambda_7$, 
(\ref{choice}), with $\Omega_{lk}$ being an orthogonal matrix in  the colour space.

For the "antisymmetric" choice (\ref{choice}), we have
\be \label{b12} b_i\equiv\frac{g}{2i}  b_{ia} \tau  ^a=
g\frac{b_i^1\lambda^2+b_i^2\lambda^5+b_i^3\lambda^7}{2i};~~~~~ b_i^a=\frac{\epsilon^{aik}n^k}{gr} \quad ( \tau_1 \equiv \lambda_2,\tau_2 \equiv \lambda_5,\tau_3 \equiv\lambda _7). \ee

\medskip For the spontaneous breakdown of the initial $SU(3)_{\rm col}$ gauge symmetry in the (\ref {break}) way, involving herewith antisymmetric Gell-Mann matrices $\lambda_2$, $\lambda_5$, $\lambda_7$ as generators of the "intermediate" $SU(2)_{\rm col}$ gauge symmetry, this BPS (Wu-Yang) monopole background takes the look (\ref {b11}) \cite{David3}.

\medskip In another aspects such Minkowskian constraint-shell QCD posesses the  in principle same "physics" that the Minkowskian YMH model with vacuum BPS monopoles quantized by Dirac, us discussed above.

In particular, if we factorize the vacuum manifold of the Minkowskian constraint-shell QCD,
$$R_{\rm QCD}= SU(2)_{\rm col}/U(1) $$
 in the (\ref{RYM}) \cite{disc} wise, this involves the Gribov "discrete" factorisation of the (\ref {fact2}) type for the "intermediate",  $ SU(2)_{\rm col}$, and residual, $ U(1)$, gauge symmetries groups.  \par 
 As regards the initial, $SU(3)_{\rm col}$, gauge  group, it is not important for us, generally speaking, what a geometrical structure ("continuous" or "discrete") has this group. 
 The only   important things are the geometries (topologies) of the "intermediate",  $ SU(2)_{\rm col}$, and residual, $ U(1)$, gauge symmetries groups.  \par 
 Supposing the Gribov "discrete" factorisation (\ref {fact2})  for $ SU(2)_{\rm col}$ and $ U(1)$ (implying the factorisation (\ref{RYM}) for the QCD vacuum manifold $ R_{\rm QCD}$), we get once again  topological rotations (\ref{rot}) (explained as a specific Josephson effect \cite{David2,rem3,Pervush3})  for the gluonic Bose condensate, involving vacuum "electric" monopoles (\ref{el.m}) and the Poincare invariant Bose condensation Hamiltonian $H_{\rm cond}$, (\ref{Hamilton}).

\medskip To  write down the Dirac equation for a quark in the BPS (Wu-Yang) monopole background (\ref{b11}), (\ref{b12}) \footnote{This is the first step in getting spectra of mesonic and baryonic 
 states. in QCD. For the detailed description of the algorithm how to do this in Minkowskian constraint-shell QCD we refer our readers to the survey \cite{Pervush2} with the numerous references therein and also to the recent work \cite{Pervush4}.}, note that 
each fermionic (quark) field may be decomposed by the complete set of the generators 
of the  Lee group $SU(2)_{col}$
(i.e. $\lambda_2, \lambda_5, \lambda_7$ in the considered case) completed by
the unit matrix $\bf 1$. This involves the following decomposition \cite{David3} of a quark field by the antisymmetric Gell-Mann matrices $\lambda_2$, $\lambda_5$, $\lambda_7$
\be \label{decompoz}
\psi_{\pm}^{\alpha,\beta} =s_{\pm} \delta^{\alpha,\beta} + v_{\pm}^j\tau_j^{\alpha,\beta},
\ee
involving some $SU(2)_{\rm col}$ isoscalar,  $s_{\pm}$, and isovector, $v_{\pm}$,
amplitudes.
 $+,-$ are spinor indices,  $\alpha,\beta$ are 
 $SU(2)_{\rm col}$ group space indices and 
 $$(\lambda_2, \lambda_5, \lambda_7)\equiv (\tau_1,\tau_2,\tau_3).$$
The mix of group and spinor indices generated by Eqs. (\ref{b11}), (\ref{b12}) for the BPS (Wu-Yang) monopole background allows  then to derive, utilising the decomposition (\ref{decompoz}), the system of differential equations in partial derivatives \cite{David3}
\be \label{Desyst} (\mp q_o+m) s_{\mp} {\mp}i(\partial_a + \frac{n_a}{r}) v^a_{\pm}=0;
\ee \be \label{Desyst1} (\mp q_o+m) v^a_{\mp} {\mp}i(\partial^a - \frac{n^a}{r}) s_{\pm}
-i\epsilon^{jab}\partial_jv_{\pm}^b =0
\ee 
(implicating the mass $m$ of a quark and its complete energy $q_0$), mathematically equivalent to the Dirac equation
\be
\label{De}
 i\gamma_0 \partial_0 \psi + \gamma_j [i\partial_j \psi+ \frac{1}{2r} \tau_a \epsilon^{ajl}n_l \psi] -m \psi= 0 \ee 
for a quark in the BPS (Wu-Yang) monopole background. 

The decomposition (\ref{decompoz}) \cite{David3} of a quark field implies that $v_{\pm}^j\tau_j^{\alpha,\beta}$ is a  three-dimensional axial vector in the colour space. 
Thus the spinor (quark) field $\psi_{\pm}^{\alpha,\beta}$ is transformed, with the "antisymmetric"
choice
$\lambda_2, \lambda_5, \lambda_7$,
by the \it reducible \rm
representation of the $SU(2)_{col}$ group that is the direct sum of the
identical representation $\bf 1$ and  three-dimensional  axial vector representation,  we denote as ${\bf 3}_{ax}$:
$${\bf 3}_{ax}\oplus {\bf 1}. $$
A  new situation, in comparison  with the
usual $SU(3)_{\rm col}$ theory in the Euclidian space $E_4$, appears in 
this case. 
That theory was worked out by Greenberg 
\cite{Greenberg},  Han and Nambu  \cite{HN,Nambu}; its goal was getting  hadronic wave 
functions  (describing bound quark states) with the correct spin-statistic connection. 
To achieve this, the \it irreducible \rm colour triplet (i.e.  three additional 
degrees of freedom of quark  colours\rm, forming the {\it polar} vector in  the $SU(3)_{\rm col}$ group space), was introduced. There was postulated that  only  colour singlets are physical
observable states. So the task of the colours confinement was outlined.\par 
Going over  to the Minkowski space in Minkowskian constraint-shell QCD quantized by Dirac and involving the (\ref {break})  breakdown of the $SU(3)_{col}$ gauge symmetry, the antisymmetric   Gell-Mann matrices $\lambda_2$, $\lambda_5$, $\lambda_7$ and BPS (Wu-Yang) physical background, allows to introduce the new, reducible,  representation of the $SU(2)_{\rm col}$ group with axial colour vector and colour scalar. \par 
In this situation the question about the physical sense of the axial colour vector $v_{\pm}^j\tau_j^{\alpha,\beta}$ is posed. \par 
For instance, it may be assumed that the axial colour vector $v_{\pm}^j\tau_j^{\alpha,\beta}$ has the form ${\bf v}_1= {\bf r} \times\bf K$, with $\bf K$ being the polar colour vector 
($SU(2)_{\rm col}$ triplet).
These quark rotary degrees of freedom corresponds to rotations of fermions together with the gluonic BPS monopole vacuum describing by  the free rotator action (\ref{rot}). The latter one is induced by  vacuum "electric" monopoles
(\ref{el.m}). These vacuum "electric" fields are, apparently, the cause of above  fermionic rotary degrees of freedom \rm (similar to rotary singlet terms in two-atomic molecules;  see e.g. \S 82 in \cite{Landau3}).  

More exactly, repeating the arguments of Ref. \cite{Pervush3}, one can "nominate" the candidature of the "interference item"
\be \label{ii}
\sim Z^a j_{Ia0},
\ee
involving the  $Z^a$ and the fermionic (quark) topologically trivial (i.e. gauge-invariant) current $j_{Ia0}^\mu=e \bar q^{I}\gamma^\mu q^{I}$, in the constraint-shell Lagrangian density of Minkowskian QCD quantized by Dirac.

\medskip
 The appearance of fermionic rotary degrees of freedom ${\bf v}_1$ in Minkowskian constraint-shell QCD  quantized by Dirac confirms indirectly the existence of the BPS monopole background in that model (coming to the Wu-Yang one \cite{Wu} at the spatial infinity).
These fermionic rotary degrees of freedom testify in favour of nontrivial topological collective vacuum dynamics proper  to the Dirac fundamental quantization \cite{Dir} of Minkowskian constraint-shell QCD (this vacuum dynamics was us described above).

\bigskip 4. This is the possibility to solve the $U(1)$ problem basing upon the Minkowskian non-Abelian YMH model with vacuum BPS monopoles quantized by Dirac. In other words, one can find the $\eta'$-meson mass near to modern experimental data.

As it was demonstrated in the recent papers  \cite{LP1,David3,LP2, Pervush2,  David1}, the way to solve the $U(1)$-problem in the Minkowskian non-Abelian Higgs model quantized by Dirac is associated with the manifest rotary properties of the apropriate physical vacuum involving YM and Higgs BPS monopole solutions.
The principal result obtained in the mentioned works regarding  solving of the $U(1)$-problem in the Minkowskian non-Abelian Higgs model quantized by Dirac is the following.

The $\eta'$-meson mass $m_{\eta'}$ proves to be inversely proportional to  $\sqrt I$, where the rotary momentum $I$ of the physical Minkowskian (YM-Higgs) vacuum is given by Eq.  \cite{David3}
$$ m_{\eta'}\sim 1/ \sqrt I,     $$
with the rotary momentum $I$ given in (\ref{I}).

More precisely,
\be
\label{mQCD}
m_{\eta'}^2 \sim \frac{C_{\eta}^2}{I V}= \frac{N_f^2}{F_{\pi}^2}\frac{\alpha_s^2<B^2>}{2\pi^3},
\ee
involving a constant $ C_{\eta}= (N_f / F_{\pi}) \sqrt {2/\pi}$, where $ F_{\pi}$ is the pionic decay constant  and $N_f$  the number of flavours in the considered Minkowskian non-Abelian Higgs model.

The explicit value (\ref{I}) of the rotary momentum $I$ of the physical Minkowskian (YM-Higgs) vacuum was substituted in  this equation for the $\eta'$-meson mass $m_{\eta'}$. 
The result (\ref{mQCD}) for the $\eta'$-meson mass $m_{\eta'}$ is given in Refs. \cite{LP1,David3,LP2, Pervush2,  David1} for the Minkowskian non-Abelian Higgs model quantized by Dirac and implemented vacuum BPS monopole solutions allows to estimate the vacuum expectation value of
  the apropriate "magnetic" field $\bf B$ (specified in that case via the Bogomol'nyi equation (\ref{Bog})) 
 \be
\label{Bav} 
 <B^2>=\frac{2\pi^3F_{\pi}^2
 m^2_{\eta '}}{N_f^2\alpha_s^2}=\frac{0.06 GeV^4}{\alpha_s^2}
 \ee 
by
using  the estimation $\alpha_{s}(q^2 \sim 0)\sim 0.24$  
\cite{David3,Bogolubskaja}. \par 

\bigskip 

The constraint-shell Abelian model (the objective of the present study) is by far simpler than the constraint-shell non-Abelian model. But there is a common point by the both these models, this is the Gauss law constraint (\ref{Gau}) resolved in terms of the Dirac variables (\ref{per.Diraca}) (where the gauge matrices $\tau^a$ turn out to the trivial unit matrix in the $U(1)$ case), {\it always gauge invariant and Poincare covariant}.

The goal of the present study is just to demonstrate this with the example of constraint-shell QED  and to generalize this constraint-shell QED (which is the topologically trivial theory) on the case of nontrivial topologies inherent in the $U(1)$ gauge group due to the natural isomorphism
\be \label{U1}
U(1)\simeq S^1
\ee
with $\pi_1 S^1={\bf Z}$.

As it is well known (see, for instance, the monographs \cite{Cheng,Ryder}), these nontrivial topologies induce the {\it Dirac monopole} (Dirac string) \cite{Dirac}, the purely gauge solution singular along the negative direction of the axis $z$ in the chosen reference frame and the {\it magnetic charge} {\bf m} satisfying the {\it Dirac quantization condition} \cite{Cheng,Ryder,Dirac}
\be \label{qq}
\frac {{\bf q m}}{4\pi}=\frac 1 2 n,\quad n\in {\bf Z},
\ee
(with $\bf q$ being the electric charge inherent in the Abelian $U(1)$ model with the unbroken gauge symmetry).

In these circumstances, we shall atempt to write down the topological Dirac variables in this Abelian $U(1)$ constraint-shall model, similar to those $\hat A_k^D$ \cite{LP1,LP2,David2} appearing in the Minkowskian non-Abelian YMH model with vacuum BPS monopoles quantized by Dirac. Such topological Dirac variables should take acount of the Dirac quantization condition (\ref{qq}) and the Dirac monopoles being presented.

\bigskip The article is organized as follows. Section 2 is devoted to the analysis of  constraint-shell QED and contains two subsections: in the first one we construct Dirac variables and performe the reduction of the QED Lagrangian in terms of Dirac variables, removing the longitudial degrees of freedom, which are unphysical. 
In the second subsection we study the Poincare covariance of the Dirac variables in  constraint-shell QED.  In Section 2 we utilize the results of the papers \cite{Pervush2,Azimov,Nguen2} and also get some new results. 

Section 3 we devote to constructing  Abelian $U(1)$ constraint-shall model involving unbroken gauge symmetry.

\section{Four-dimensional constraint-shall QED.}
\subsection{Constructing Dirac variables in four-dimensional constraint-shall QED.}
Let us consider the standard QED action \cite{Pervush2}
 \begin{eqnarray}
\label{QEDa}  
W[A,\psi,\bar \psi] &=& \int d x \,\,\, \Bigl[ -\frac{1}{4} ( F_{\mu\nu}  )^{2} + \bar {\psi} ( i \,\, \rlap/\nabla (A) - m^{0} ) \psi  \Bigr],  
\end{eqnarray}
with
with
\begin{eqnarray}
\label{vw}
 \nabla_{\mu}(A) &=& \partial_{\mu} -i e {A}_{\mu}, \quad  \rlap/\nabla = \nabla_\mu \cdot \gamma^\mu;  \nonumber \\  \,\,\,\,\,F_{\mu\nu} &=&  \partial_{\mu}A_{\nu} - \partial_{\nu}A_{\mu}.
 \end{eqnarray} 
This action  contains gauge fields more than the number of independent degrees of freedom. \par 
The action  (\ref{QEDa}) is invariant under the gauge transformations  
\be
\label{gauge}  
 A_{\mu}^\Lambda=A_{\mu}+ \partial_{\mu}\Lambda, \quad \psi^\Lambda=
 \exp[ie\Lambda]\psi,   \quad   \partial _\mu \partial^\mu\Lambda =0.  \ee 
There may be shown (see \S 19.5 in \cite{A.I.}) that the gauge transformations (\ref{gauge}) can be represented in the canonical form 
$$
A_{\mu}' =UA_{\mu} U^{-1} ,$$ $$ \psi '= U\psi U^{-1},$$  
\be
\label{gauge'} 
\bar \psi '= U \bar \psi U^{-1} \ee  with  a unitary operator $U$: 
$$ U U^{-1} =I.$$ 
Issuing from the gauge transformations (\ref{gauge}), one can  represent the operator $U$  in the form  
\be
\label{unitary}
 U=e^{iF}, \ee 
with $F$ being a Hermitian  operator, $F=F^\dagger$, having the look 
 \be\label{F}
F= \int \{ \Lambda(x) \frac{ \partial \chi}{ \partial t}- \frac{ \partial  \Lambda(x) }{ \partial t}  \chi\} d^3x, \ee  
with
$$ \chi = \frac{ \partial A_\mu (x)}{ \partial x_\mu}.$$ 
The gauge transformations (\ref{gauge'}) form, obviously, the Abelian group $U(1)$. 

For 
infinitesimal  gauge transformations (\ref{gauge}),  (\ref{gauge'}) the function $\Lambda(x)$ also will be 
infinitesimal.  
On the other  hand, in this case $$ U\approx I+iF,$$
  and Eqs. (\ref{gauge'}) acquire the look 
$$ A_{\mu}' \approx A_{\mu} + i[F, A_{\mu} ],$$ 
$$ \psi '\approx\psi + i[F, \psi],$$
  \be\label{gauge''} \bar \psi '\approx\bar \psi  + i[F, \bar\psi]. 
\ee 
The comparison of the  gauge transformations (\ref{gauge}) and  (\ref{gauge''})  results  the relations
$$ i[F, A_{\mu} ] =  \frac{ \partial \Lambda(x)}{\partial x_{\mu}},$$ 
$$  i[F, \psi] = i e  \Lambda(x) \psi,$$  
\be
\label{rela} 
i[F, \bar\psi]= -  i e  \Lambda(x)\bar\psi. \ee
Because of the infinitesimal nature of the considered gauge transformations,
one can then identify the  functions $\Lambda(x)$ and $F(x)$. \par 
 Note that the QCD Lagrangian (\ref{QEDa}) remains invariant  under the transformations  (\ref{gauge'}) combined with (\ref{gauge}) \cite{Bogolubskaja}:
 \be
\label{3.9} 
{\cal L} (\hat A^g) ={\cal L} (\hat A) \quad {\rm at} \quad \hat A^g = g(\hat A+\partial) g ^{-1}\equiv \hat A_\mu \to \hat A_\mu -i e\partial_\mu \Lambda(x).  \ee 
 
Herewith 
\be
\label{gm}
g\equiv \exp[ie\Lambda(x)].
\ee
In Eq . (\ref{3.9}) the record $\hat A$ stands for denoting 
\be \label{edinizi} \hat A_\mu = i \frac{e}{\hbar c} A_\mu,  
\ee
where the correct account of the elementary charge $e$ and Planck constant $\hbar $ in QED is taken. 

In this case the "coupling constant" $e/(\hbar c)$ would  also enter gauge transformations (\ref{gm}). Herewith the function $\Lambda(x)$ is chosen in such a wise that exponential multipliers in (\ref{gm}) become dimensionless.

As one acts usually in QFT, we shall apply the Planck system of units, where $\hbar =c =1$ is set, in the majority of formulas in the present study. 
Simultaneously, sometimes we shall write down explicitly these constants in the cases when their role is important for understanding properties of physical models we represent in our work.\par  

\bigskip 

Let us now suppose that the invariance of QED  with respect to the  gauge transformations (\ref{gauge}),  (\ref{gauge''}) \cite{Pervush2,A.I.} allows  to remove a one field degree  of freedom with the aid of an arbitrary gauge
 \be \label{gf}
 F(A_{\mu})=0, \quad F(A^{u}_{\mu})=M_F u \not= 0,  \ee 
where the second equation means that the given gauge unambiguously fixes the field $A$.\par 
In general,  to construct QED as a quantum-field theory obeyed the usual Feynman rules, one would always fix a certain  gauge: say, $f_i=0$; and this fact has several  consequences \cite{Bogolubskaja}. \par 
1. The explicit solution to the constraint  $f_i=0$ gives the definite class of  physical \rm
(gauge invariant) variables, functionals on initial  gauge \rm fields $A_i$ \cite{Pervush3}.\par 
The most important patterns of such functionals are the transverse and longitudinal  physical  fields.  
 Maxwell electrodynamics   gives us an example of transverse  physical  fields. There are electric and magnetic tensions associated with the plane electromagnetic wave. 
 To get in this case the D'alembert equation (see e.g. \S 46 in \cite{Landau2}) \be
\label{D'alambert}
 \Delta {\bf A}- \frac{\partial ^2{\bf A}}{\partial t^2}=0,
\ee 
one would utilize  two constraints in the  QED Hamiltonian formalism. \par
 The first of those constraints
is the  of the  secondary constraint 
\be\label{3.12} A_0\equiv \phi=0,\ee
referred to as the Weyl gauge in modern physical literatyre. This removal of the temporal component of a four-potential $A$ in QED is quite justified due to the trivial canonical momentum $\partial L/\partial \dot A_0=0$ inherent in this particular model \cite{Gitman} (see our 
 in Introduction).\par 
The second constraint  one utilizes in  Maxwell electrodynamics  at deriving plane wave solutions is the  combination \rm of the secondary  constraint (\ref{3.12}) and the other secondary constraint, us familiar already   the Gauss law constraint (\ref{Gau}). Note here the remarkable property  of the Gauss law constraint (\ref{Gau}): {\it it is simultaneously the motion equation and the secondary constraint} \cite{Gitman} (that is correctly for any particular theory). 

The  more detailed look  of the Gauss law constraint in QED will be us cited below; now we only note
that  the combination of the  both constraints, (\ref{3.12}) and (\ref{Gau}) comes, indeed, to the 
\it radiation \rm (\it Coulomb\rm) \it gauge \rm \footnote{In 4-dimensional QED, if fermions are absent, the Gauss law constraint (\ref{Gau}) is expressed as \cite{Gitman}
$$ \bigtriangleup A_0+\partial_0(\partial_i A^i)=0.$$
}
\be\label{Kulon11} {\rm div} ~ {\bf A}=0, \ee  
that is the spatial part of the \it Lorentz gauge \rm
\be
\label{Kulon} 
\partial_\mu A ^\mu =0.  \ee 
The latter one, in turns, comes, for the electric tension ${\bf E}$,
 to  the Maxwell equation 
 $$ {\rm div}~ {\bf E} =- {\rm div} ~ \frac{\partial {\bf A}}{\partial t} = -\frac{\partial }{\partial t}~{\rm div} ~ {\bf A}=0$$ 
when  electromagnetic currents are absent (i.e.  in the lowest order  of the perturbation theory) and the Weyl gauge (\ref{3.12}) is taken. \par
 Note  that the  latter formula is the specific expression for the "Maxwell"
Gauss law constraint when electromagnetic currents are absent and the Weyl gauge (\ref{3.12}) is taken (cf. (15.12) in  \cite{Gitman}).

\medskip 
  \be
\label{naprjag} 
{\bf E} =-   \frac{\partial {\bf A}}{\partial t} 
\ee 
is an example of gauge invariant physical  \it local functionals \rm of  gauge fields.

\medskip The Coulomb gauge 
$$ {\rm div} ~ {\bf A}={\rm div} ~{\bf E} =0$$  
implies that the four-potential  $A$ and electric tension ${\bf E}$ are orthogonal to the momentum
$p_i=-i\partial_i$, i.e.  transverse\rm. \par 
Indeed there may be  shown \cite{Rohrlich} that the secondary constraint (\ref{3.12}) follows directly from the  Lorentz gauge (\ref {Kulon}) since the four-vector of momentum, $p$, is a null vector (i.e. $\vert {\bf p}\vert = 0$) for an electromagnetic field. \par 
Really,
one may rewrite Eq. (\ref {Kulon}) as
\be \label {Kulon1}p_\mu A^\mu=0;\ee
therefore  
\be \label {A00} A_0= {\bf p \cdot A}/ p_0\ee 
in the Minkowskian signature $(+,-,-,-)$. 

Thus  the temporal component of $A^\mu$ is
eliminated  by  cancelling  the longitudinal  space-like  component of the  four-potential\rm, that is
\be
 \label {sp-like}
{\bf p \cdot A}=0.
\ee 
At the quantum level,  the Lorentz gauge (\ref {Kulon}),  (\ref {Kulon1}) come to the (\it weak\rm) condition
\be \label{weak}
p\cdot A \vert \phi >=0\ee
imposed onto the physical state vectors of the Hilbert-Fock space $\cal H$ of the second 
quantization. \par
 Thus  eliminating   temporal component $A^0$ with   cancelling   the longitudinal  space-like  component ${\bf p}\cdot {\bf A }$ of a  four-potential $A$ implies that the of negative norm states (\it ghosts\rm) are absent in the Lorentz gauge (\ref {Kulon}),  (\ref {Kulon1}) in Gauss-shell electrodynamics   (accompanied by  the  null energy-momentum four-vector $ p^2 =0$ \footnote{In the case of massive vector bosons with the spin 1, described in the Proca model, $ p^2 =M$, with $M$ being the mass of the spin 1 vector boson, and also in this case \cite{Ryder} ${\bf p}\cdot {\bf A }\neq 0$: massive vector bosons with the spin 1 always possess longitudinal  space-like  components. \par
Thus \cite{Rohrlich}  for massive vector bosons with the spin 1 eliminating  their temporal components $A^0$ is possible in the only case when $\vert {\bf p}\vert =0$, i.e. in a c.m. reference frame in the Minkowski space-time.}. \par \medskip
Another example of transverse  physical fields are \it Dirac variables \rm \cite{Pervush2}, the topic of our discussion in the present study.

 \medskip 2.  The change of the gauge for field functionals ($A^{f_1}\to A^{f_2}$) is fulfilled by  substituting \cite{FP1,Abers}
 \be
\label{change of the gauge}
 A^{f_2}[A^{f_1}] = V[A^{f_1}] (A^{f_1}+\partial) V^{-1} [A^{f_1}]; \quad \psi ^{f_2}= V[A^{f_1}] \psi ^{f_1}. 
\ee
Herewith there may be shown that all the Green functions  
are invariant with respect to changes of  gauges (\ref{change of the gauge}); for example:
\be\label{3.17}
<\psi ^{f_2}\cdots \bar \psi ^{f_2}>\equiv <V[A^{f_1}] \psi ^{f_1}\cdots \bar \psi ^{f_1}V^{-1} [A^{f_1}]> 
\ee 
(in theories without  anomalies\rm).

\bigskip To proceed further, note that classical  equations of a gauge model are split into the {\it constraints}, which relate initial data for
spatial components of the fields involved in a (gauge) model to initial data of their temporal components, and the {\it equations of motion}: to solve these, it is necessary
to measure initial data \cite{LP2}. 

The both classes can intersect, for instance, in the case of the Gauss law constraint (\ref{Gau}): as it was discussed above, it is simultaneously the motion equation and the secondary constraint \cite{Gitman}.

This fact plays a very important role in the quantization of  particular theories \cite{Gitman}, in particular, in the quantization of constraint-shell QCD, the topic of the present study.

\medskip Above we have demonstrated that the Gauss law constraint (\ref{Gau}) is reduced to Eq. (\ref{Kulon11}) in  pure "Maxwell" electrodynamics, i.e. when fermionic electromagnetic currents $j$ are absent. Now let us assume that latter are "switched on". In this case the Gauss law constraint (\ref{Gau}) acquires the look
\be 
\label{var1} 
 \frac{\delta W}{\delta A_0}=0  ~\Rightarrow~~  \Delta A_0= \partial_i  \partial_0 A_i +j_0; \quad ~~~~\Delta =\partial_i \partial_i,~j_{\mu}=e\bar\psi\gamma_{\mu}  \psi   \ee 
(we refer,  in the present work, the Latin indices to the spatial field components).

On the other hand, the set of equations of motion in such QED looks as \cite{Pervush2}
\be 
\label{var2}  
\frac{\delta W}{\delta A_k}=0~  ~\Rightarrow~~  \partial_0^2 A_k- \partial_k \partial_0 A_0-(\delta_{ki}\Delta-  \partial_k \partial_i ) A_i=j_k,  
\ee 
 \be
 \label{var3} 
 \frac{\delta W}{\delta \psi}=0~  ~\Rightarrow~~\bar \psi ( i \,\, \rlap/\nabla (A) + m^{0} )=0,  \ee  
$$  \frac{\delta W}{\delta \bar \psi}=0~  ~\Rightarrow~~( i \,\, \rlap/\nabla (A) - m^{0} )\psi=0  $$
(here $\rlap/\nabla=\nabla \cdot \gamma$).
 
  \medskip
The problem of the canonical quantization encounters  the nondynamical status of  temporal fields components $A_0=A_\mu  \eta_\mu$ (with $\eta_\mu$ being the chosen reference frame) \footnote{There can be given the general definition of a reference frame \cite{LP2} as a set of physical instruments for measuring initial data in a physical theory. \par 
In this context one speak about {\it inertial}  reference frames in special relativity (SR). This means that the given coordinate basis is connected with a
heavy physical body  moving  without influences of any  external
forces \cite{Dub}

Customary,  inertial  reference frames in the Minkowskian space-time are
associated with the unit time axis  
$$\eta_\mu=
 (\frac{1}{\sqrt{1-\vec v^2}},\frac{\vec v}{\sqrt{1-\vec v^2}}),$$ 
with $\vec v$ being the velocity of a physical body.

The frame of reference $\eta^{\rm 0}_\mu=(1,0,0,0)$ with $\vec v=0$
 is called the comoving frame (in the present study we shall  apply often the term ``rest reference frame`` to such reference frames). In the present study we  shall utilize very often  comoving  frames, in which  a body is immovable. 
 
\medskip In this terminology, one can define \it relativistic transformations \rm as those that change initial data (these can be written down \cite{Dub} as $L_{\mu\nu}\eta^{\rm 0}_\mu=\eta_\mu$) and  \it gauge transformations \rm as those that  do not affect the readings of   instruments and are associated with gauges of physical fields.}.  
The non-dynamic status of $A_0$ is not compatible with the
 quantization of this component of an electromagnetic field as fixing  $A_0$ (via the Gauss law constraint (\ref{var1})) and  its zero momentum  
\be 
\label{zero momentum} 
E_0=\partial {\cal L}/\partial (\partial_0
 A_0)=0
\ee 
contradict  the commutation relations and Heisenberg uncertainty principle. \par 
Besides that,  temporal components $A_0$ of gauge fields result the of negative norm states (ghosts) at the second  quantization of any physical theory  \footnote{The Gauss law constraint (\ref{var1}) acquires in QED the look \cite{Landau2} ${\rm div}~ {\bf E}= j_0$ in the fermions present. 

The Gauss law constraint (\ref{var1}) is the  secondary constrain  obtained \cite{Prohorov} at the commutation relation between the QED Hamiltonian $\hat H$ and the canonical momentum $E_0$: 
$$[E_0, \hat H]= i({\rm div}~ {\bf E}-j_0)\approx 0;$$  
at the commutation relation between the QED Hamiltonian $\hat H$ and the canonical momentum $E_0$: 
$$[E_0, \hat H]= i({\rm div}~ {\bf E}-j_0)\approx 0;$$  while the equal to zero  canonical momentum $E_0 $ is the \it the primary constraint \rm in the QED Hamiltonian formalism. On the other hand, since $[E_0,{\rm div}~ {\bf E}-j_0]= 0$,  both the mentioned constraints belong to the first class of constraints.
}.

To keep the quantum principles, Dirac excluded  temporal components of gauge fields  using the Gauss law constraint~(\ref{var1}): herewith the explicit solution to the Gauss law constraint is
explicit solution to the Gauss law constraint is
 \be 
\label{var1s} 
 A_0(t,x)= a_0[A]+\frac{1}{\Delta}j_0(t,x),  \ee  
where  
\be 
\label{var2s}   a_0[A]=\frac{1}{\Delta}  \partial_i  \partial_0 A_i(t,x) 
\ee  
associates the initial data of $A_0(t_0,x)$ to the set of
  initial data of the  longitudinal component $\partial_i \partial_0 A_i(t,y)$  and the (fermionic) current $j_0(t,y)$ in the whole space. 

 Here  
 \be 
\label{coulomb} 
 \frac{1}{\Delta}f(x)=-\frac{1}{4\pi}\int d^3y\frac{f(y)}{|\bf{x}-{\bf y}|}
 \ee
 is the Coulomb kernel of the apropriate nonlocal distribution. 

As we remember from mathematical physics (see e.g. p. 203 in \cite {V.S.Vladimirov}),  the \it fundamental solution \rm to the \it Laplace equation \rm 
\be 
\label{Laplace}
 \Delta {\cal E}_3 =\delta (x)  \ee 
is
\be 
\label{fund}
 {\cal E}_3 = -\frac {1}{4\pi x}. 
\ee 
Just this specifies the action of the operator $\Delta^{-1}$, (\ref{coulomb}), on a continuous function $f(x)$.\par
Taking into account (\ref {coulomb}),  Eq. (\ref{var1s}) may be rewritten in the integral form as \cite{Azimov}
\be 
\label{var1si}
A_0(t,x)= -\frac{1}{4\pi}\int \frac{d^3y}{|\bf{x}-{\bf y}|} (\partial_i  \partial_0 A_i(t,y)+j_0(t,y)). \ee 
One can  substitute the solution (\ref{var1s}) into the equation (\ref{var2}) for spatial  components:  
\begin{eqnarray} 
 \label{vartr}  
\frac{\delta W}{\delta A_i}\;\Bigl\vert_{ \frac{\delta W}{\delta
 A_0}\;=\;0}~\Rightarrow~[\delta_{ik}-\partial_i\frac{1}{\Delta}\partial_k]
 (\partial_0^2-\Delta) A_k=j_i-\partial_i\frac{1}{\Delta}\partial_0
 j_0. 
\end{eqnarray}
 We  see that the constraint-shell equations of motion (\ref{vartr}) contain only two transverse physical variables that are, indeed, gauge invariant functionals:
  \be 
\label{**}  
A^*_i(t,{\bf x}) =  [\delta_{ik}-\partial_i\frac{1}{\Delta}\partial_k]  A_k.  \ee
(one can  make sure directly that  variables
(\ref {**}) are  gauge invariant functionals by comparing the latter formula  with the gauge transformations~(\ref{gauge})). 

Dirac rewrote these gauge invariant variables with the aid of  the gauge transformations  \cite{Pervush2,Azimov}
$$  \sum_{a=1,2} e_k^aA_a^D\equiv A_{k}^{D}[A]=  v[A] ( A_{k} + i \frac{1}{e} \partial_{k} )   v[A] ^{-1},  $$ 
 \be 
\label{vak} 
 \psi^{D}[A,\psi] = v[A] \psi,   \ee   
 where the gauge factor $v[A]$ \cite{Dir} was defined as
  \be
\label{va1}  
v[A] =  \exp \bigl\{-ie\int_{t_0}^{t} dt' a_0 (t')\bigr\}. 
 \ee
It is obvious \cite{Nguen2} that gauge invariant variables $A_k^D$, $\psi^{D}$ belong to the {\it Heisenberg representation} for quantum-field operators.
 
\medskip Let us now  use  the gauge transformation (\ref{gauge}) for   temporal
components $a_0$ of electromagnetic fields:  
 \be
\label{vgauge} 
 a^{\Lambda}_0=a_{0}+\partial_0 \Lambda~\Rightarrow~  v[A^{\Lambda}]= \exp[ie\Lambda(t_0,{\bf x})]v[A]\exp[-ie\Lambda(t,{\bf x})]. 
 \ee 
But it is the same transformations law that  (\ref{gauge'}) \cite{A.I.} upon identifying the functions $\Lambda$ and $F$ for infinitesimal gauge transformations. 

Comparing then Eqs.  (\ref{vgauge})  and (\ref{3.9}), we draw the conclusion that  should functionals \rm (\ref{vak}) be, indeed, gauge invariant, it is sufficient that Dirac gauge factors transformed\rm,  (\ref{vgauge}), cancel the transformation law \rm (\ref{3.9}) \footnote{Herewith the stationary  matrices $\exp[ie\Lambda(t_0,{\bf x})]$ would be included in the apropriate gauge transformations (\ref{3.9}) to cancel entirely this transformation law for multipliers $ v[A] $.}.

Thus \cite{Azimov} one would claim
\be
\label{canz}
v[A^{\Lambda}]= v[A] g^{-1}.
\ee
In this case the simple computation \cite{Pervush3,Azimov}
\be \label{proverca}
 A_k^D[A^{\Lambda}]= v[A]~ g^{-1} g(A_k+ i \frac{1}{e} \partial_k) g^{-1} g ~v[A] ^{-1} = A_k^D \ee
gives the  way to verify exactly the  gauge invariance of nonlocal functionals \rm (\ref{vak}).  

We shall call  functionals \rm (\ref{vak}) (or, equivalent, (\ref{**})) \it the Dirac variables \rm (for instance, in the terminology \cite{Pervush2}).

The fact of gauge invariance of Dirac variables (\ref{vak}) is very remarkable. Indeed, the conception of gauge invariant Dirac variables as representative for four potentials in a  gauge model is a new way therein in comparison with gauge fixing method (applied usually in QED, Yang-Mills theory, QCD and so on). This is the main advantage of this concept, on the author opinion.

\bigskip

 We learn from (\ref{vgauge}) that the initial data of  the gauge invariant Dirac variables~(\ref{vak}) are  degenerated with respect to a  stationary phase \rm
 \be
\label{faza} 
\exp[ie\Lambda(t_0,{\bf x})]\equiv \exp[ie\hat \Phi_0({\bf x})] .
\ee 
The Dirac variables (\ref{vak}),  as  functionals of initial data,  satisfy the
identity
\be
\label{gc2}  
\partial_0 \left(\partial_i A^D_i(t,{\bf x}) \right) \equiv 0
 \ee 
in the purely electromagnetic theory without electronic currents (i.e. if $j_0\equiv 0$). 
This identity also may be checked directly (issuing from Eqs. (\ref{vak}), (\ref{va1}), which result Eq.  (\ref{**}) if decompose the exponent in (\ref{va1}); in turn, the functionals (\ref{**})  satisfy (\ref{gc2})). 
Eq. (\ref{gc2})  implies that Dirac variables~(\ref{vak}) are  transverse functionals of   gauge fields\rm.  

The identity (\ref{gc2}) is obtained formally from the Gauss law constraint  
\be 
\label{Gausel} 
 \frac{\delta W}{\delta A_0}=0  ~\Rightarrow~~  \Delta A_0= \partial_i  \partial_0 A_i 
\ee 
in the {\it purely} electromagnetic theory without electronic currents \footnote{It is correctly in the lowest order of the perturbation theory by $e^2/(\hbar c)$ at substituting $A_0^D=0$ in the latter equation, i.e.  at the Dirac removal \cite{Dir} of 
nondynamical temporal field components from the Gauss law constraint  and the apropriate Lagrangian density.}.

\medskip The Lagrangian density of the purely electromagnetic theory without electronic currents has the look \cite{Gitman, Nguen2}
\be \label{LED}
{\cal L}_{\rm ed} = \frac{1}{2}F_{0i}^2-\frac{1}{2} B^2_i;
\ee
$$ F_{0i}=\partial_0 A_i-\partial_i A_0; \quad B_i= \epsilon_{ijk}\partial^j A^k =\frac{1}{2}\epsilon_{ijk} F^{jk}. $$
The Gauss law constraint  (\ref{Gausel}) permits in this case the particular solution
\be \label{var2s1}
A_0 = \frac{1}{\Delta} \partial_j  \partial_0 A_j\equiv a_0.
\ee
Substituting this solution for $A_0$ in Eq. for the electric field $F_{0i}$, we get \cite{Nguen2}
\be \label{electra}
F_{0i}= \dot A_i-\partial_i A_0=\dot A_i-\partial_i(\frac{1}{\Delta}\partial_j\dot A^j)= \delta_{ij}^D\dot A^j=\dot A_i^D \ee due to (\ref {**}), (\ref{vak}). 
Herewith the operator
\be \label{proektir}
\delta_{ij}^D\equiv (\delta_{ij}- \partial_i (\frac{1}{\Delta} \partial_j));  \quad (\delta_{ij}^D)^2=\delta_{ij}^D; 
\ee
projects out the vector ${\bf A}=(A_1,A_2,A_3)$ into the plane perpendicular to the direction of propagation of the given electromagnetic wave. 

As a result, the Lagrangian density of the purely electromagnetic theory becomes a gauge invariant expression involving only two nonlocal transverse field components $ A_i^D$ ($i=1,2$): (\ref {**}), (\ref{vak}), that are, in turn, gauge invariant under the transformations (\ref{3.9}):
\be \label{LEDt}
{\cal L}_{\rm ed}(x)= \frac{1}{2} (\dot A_i^D)^2 -\frac{1}{4}F_{ij}^2.
\ee
Mathematically, all the said is equivalent to setting in zero temporal field components $A_0$: in the Gauss law constraint  (\ref{Gausel}) as well as in the Lagrangian density (\ref{LEDt}).

\medskip
In general, when one removes  temporal components of  Abelian gauge fields (this is also true 
for non-Abelian fields), the apropriate Dirac variables becomes equal to zero \cite{LP1,David2,Azimov}:  
\be \label{udalenie1} v[A](\hat a_0(t,{\bf x})+\partial_0)v^{-1}[A]=0; \quad \hat a_0=ieA_0.
\ee 
One  can treat  latter Eq. as the equation for specifying  Dirac matrices $v[A]$ and  alone these matrices,  (\ref{va1}), as solutions to this equation.\par
As one performs the gauge transformations (\ref{udalenie1}) for temporal components of  Abelian fields, he automatically turns spatial components of these fields into transverse and gauge invariant Dirac variables (\ref {vak}) satisfying the condition (\ref {gc2}) \cite{Pervush2}.

\bigskip Now let us again return to the QED model involving fermionic currents.
In this case, as we have already ascertained, the Gauss law constraint  (\ref{var1}) permits the particular solution (\ref {var1s}). 

The  Lagrangian density of QED has the standard look (\ref{QEDa}) supplemented by the currents interaction item (see e.g. \S 17.2 in \cite{A.I.})
\be \label{vzaimod}
{\cal L}_I= j_\mu(x) \cdot A_\mu(x).  
\ee
The QED Lagrangian density (\ref{QEDa}) supplemented by the currents interaction item (\ref{vzaimod}) can be rewritten in terms of transverse and gauge invariant fields,  Dirac variables (\ref {**}), (\ref{vak}), upon substituting $A_0$, (\ref {var1s}), in this Lagrangian density. 
This results \cite{ Azimov,Nguen2}
\be \label{QEDt}
{\cal L}(x)= \frac{1}{2}F^2_{0i}(A^D)- \frac{1}{4}F_{ij}^2-j_i A_i^D + j_0 \frac{1}{\Delta} j_0+ j_0\frac{1}{\Delta} \partial_0\partial_i A_i+\bar \psi \{ i\gamma_\mu [\partial_\mu+ie \partial_\mu(\frac{1}{\Delta}\partial_i A_i)-m]\}\psi;
\ee 
$$ F_{0i}(A^D)= \dot A_i^D-\partial_i A_0^T,  \quad A_0^T=\frac{1}{\Delta} j_0(x).$$
 The electric field $\partial_i A_0^T $, entering the Lagrangian density (\ref{QEDt}),
   satisfies the Poisson equation
\be \label{puas}
\Delta A_0^T=j_0(x).
\ee
Due to the Dirac removal (\ref{udalenie1}) \cite{LP1,Dir,David2} of  temporal components of gauge fields, the fifth item in (\ref {QEDt}) would, indeed, vanish upon total going over to Dirac variables (involving the gauge (\ref {gc2})). 

In this case, upon ruling out  the surface items 
$$ \bar \psi \{-e\gamma_\mu \partial_\mu(\frac{1}{\Delta}\partial_i A_i)\}\psi,$$
  $\sim \dot A_i^D \partial_i A_0^T $ and $(\partial_i A_0^T)^2$
from the Lagrangian density (\ref{QEDt}) and replacing fermionic fields by Dirac variables $\psi^D$, $\bar \psi^D$: (\ref {vak}) (just as this was done for gauge fields $A$ in  (\ref{QEDt})), it effectively acquires the gauge invariant look \cite{Nguen2}
\be \label{QEDt1}
{\cal L}^D(x)=\frac{1}{2} (\dot A_i^D)^2-\frac{1}{4}F_{ij}^2- j_i^D A_i^D+ \frac{1}{2} j_0^D \frac{1}{\Delta} j_0^D + \bar \psi^D [i\gamma_\mu \partial_\mu-m] \psi^D, 
\ee
where the fermionic current $j^D$ is written down in terms of Dirac variables $\psi^D$, $\bar \psi^D$.

\medskip The gauge invariant Lagrangian density (\ref{QEDt1}), written down in terms of Dirac variables $A^D$, $\psi^D$, $\bar \psi^D$, (\ref {vak}), describes correctly the \it equivalent unconstrained system \rm  (EUS)  \cite{Pervush2} for QED  on the surface of the Gauss law constraint (\ref {var1}).

The apropriate Gauss law \it constraint-shell action \rm \cite{Pervush2}, describing this  EUS , can be written down as
\be \label{EQUS}
W^*=W|_{\delta W/{\delta A_0}=0}=\int d^4x {\cal L}^D(x).
\ee
To combine the nonlocal physical variables $A^D$  and variation principle formulated for these nonlocal
fields, one would consider the effective action \cite{Gitman,Pervush2} 
\be \label{efac}  W_{\rm eff}=W^* +\int d^4x \lambda_L (x)\partial_i A_i^D ~~~ (i=1,2).
\ee
with $\lambda_L (x)$ being a Lagrange multiplier.

From the gauge invariant Lagrangian density (\ref{QEDt1}) one reads the equations of motions \cite{Nguen2} \footnote{Following Dirac \cite{Dir}, we change herewith the order  constraining and varying at analysis of Gauss law  constraint-shell QCD.
The constraint-shell action (\ref{EQUS}) is got in the rest reference frame $\eta_\mu=(1,0,0,0)$.}
\be \label{Dalam}
\frac{\delta W_{\rm eff}}{\delta A^D_i}=0 \Longrightarrow \square A_k^D(x)= \delta_{ki}^D j_i^D(x)~~(i,j,k=1,2);
\ee
\be \label{Direq}
\frac{\delta W^*}{\delta \bar \psi^D}=0 \Longrightarrow   (i\gamma_\mu \partial_\mu-m) \psi^D(x)= -e \gamma_i \psi^D(x) A_i^D(x)+ \frac{1}{2} \gamma_0 \{\psi^D(x), \frac{1}{\Delta} j_0^D(x)\}, \ee
with the projection operator $\delta_{ki}^D $ given in (\ref {proektir}). 

It  projects effectively QED into the time-like surface of three-vectors $A^D\equiv(A_0^T, A_1^D, A_2^D)$ swept by the transverse components $ A_1^D$, $ A_2^D$ of an "four-potential" in the Minkowski space (that are, indeed, the gauge invariant Dirac variables (\ref {**}), (\ref{vak})) and  the temporal field component $ A_0^T $ specified in (\ref {QEDt}), (\ref{puas}) and then rewritten in terms of   gauge invariant fermionic currents $j^D$.

As a result, the temporal field component $ A_0^T $  becomes indeed "transverse" since the fermionic currents $j^D_\mu$ are such (because of their explicit look $e~\bar \psi^D\gamma_\mu\psi^D $). 

\medskip The  D'alembert equation (\ref{Dalam}) has the standard look \cite{Landau2} describing plane electromagnetic waves involving transverse polarizations of electric and magnetic tensions. 

By analogy with ordinary  Maxwell electrodynamics (see e.g., \S62 in \cite{Landau2}), the solution to the  D'alembert equation (\ref{Dalam}) can be represented in the shape of an \it retarding potential \rm written down in terms of transverse currents $j^D$:
\be \label{retard}
{\bf A}^D(x)=\frac{1}{c} \int \frac{j^D_{t-R/c}}{R}d^3x +{\bf A}^D_{(0)}(x), 
\ee
with ${\bf A}^D_{(0)}$ being the solution to the  homogeneous equation
\be \label{Dalamh}
\square ~A_k^D=0
\ee
and $R$ being the distance between the origin of coordinates and the observation point \footnote{At the "Feynman" level, the solution (\ref{retard}) to the D'alembert equation (\ref{Dalam}) describes correctly the interaction of two fermionic currents (the four-fermionic interaction), that is the  of the second order process in the perturbation theory.

Formally (see \S 32.1 in \cite{A.I.}), such interaction of two fermionic currents has the look
$$ {\bf S}^{(2)} =-\frac 1 2 \int T[j_\mu (x)j_\nu (x')] T[A_\mu (x)A_\nu (x')]d^4x d^4 x'=-\frac 1 2 \int j_\mu (x)D_c(x-x')j_\nu (x') d^4x d^4 x',$$
with 
$$ <0\vert T[A_\mu (x)A_\nu (x')]\vert 0> \equiv \delta_{\mu \nu}D_c(x-x')$$ being the photonic cause function (if physical photons are absent in the initial and final states; it is now just the case).

At these circumstances,  the matrix element $<f\vert{\bf S}^{(2)}\vert i>$ for the four-fermionic interaction can be represented in the shape of retarding potentials (where now we write down our equations in terms of Dirac variables, cf. \S 32.2 in \cite{A.I.}):
$$ <f\vert{\bf S}^{(2)}\vert i>=i\int dt \int j^D_1({\bf r}_2,t) A^D_\mu({\bf r}_2,t)d^3x_2=-e\int \bar \psi^D(x)\hat A(x) \psi^D(x) d^4x$$
with 
$$A_\mu ({\bf r},t)=\frac 1{4\pi} \int j^D_\mu ({\bf r}') \frac {e^{-i\omega t+i\omega \vert {\bf r}-{\bf r}'\vert}}{\vert {\bf r}-{\bf r}'\vert} d^3x'$$
and $\omega$ being the energy loss along two fermionic lines: incoming and outcoming, of the vertex in the apropriate second order Feynman diagram
(see Fig 32.1 in \cite{A.I.}).}. 

Thus the D'alembert equation (\ref{Dalam}) is on-shell of physical (transverse) photons, resulting the massless  photon propagator \cite{Nguen2}
\be \label{fpr}
D_{ij}^D(q)= \frac{1}{q^2+i\epsilon} (\delta_{ij}-q_i \frac{1}{{\bf q}^2} q_j) ~~ (i,j=1,2); \quad D_{il} q^l=0.  
\ee 
\medskip The second equation of motion,  (\ref {Direq}), in the system of motion equations (\ref{Dalam}), (\ref {Direq}) differs from the ordinary Dirac equation in QED by the second item on the right hand side of (\ref {Direq}).  It describes the Coulomb instantaneous interaction of fermionic currents remaining upon eliminating  $a_0$ via 
(\ref {udalenie1}) \cite{LP2,David2}.

Thus the solution to the system of  equations of motion (\ref{Dalam}), (\ref {Direq}) is a time-like three-vector  $A^D=(A_0^T, {\bf A}_i^D)$ ($i=1,2$) with the temporal component 
$ A_0^T $ being the solution to the Poisson equation 
\be \label{dub} A_0^T=\frac{1}{\Delta} j_0^D(x),\ee 
coinciding mathematically with (\ref{puas}), and the transverse spatial components ${\bf A}_i^D $ given in 
(\ref {retard}) \cite{Landau2}. The latter ones are gauge invariant retarding potentials. 

In turn \cite{Dub}, Eq. (\ref{dub}) is nothing else but the Gauss law constraint written down in terms of the fermionic Dirac variable $j_0^D(x)$. Note that it is  the effect of the manifest presence of fermionic currents and charges in (constraint-shell) QCD. In the ``pure'' electrodynamics (\ref{LED}) (with its constraint-shell reduction (\ref{LEDt})) \cite{Nguen2}, the right-hand side of Eq. (\ref{dub}) should vanish, and it becomes  homogeneous Laplace  equation \footnote{In Section 3 we shall once again give arguments in favour of
impossibility to remove the temporal component of a four-potential and, consequently, the apropriate electric tension $\bf E$ when an electric charge/current is present in the considered theory.
}. 

From the physical standpoint \cite{Dir,Dub}, all this is equivalent to the removal of the $\partial_0\partial_k A^k$ item in the Gauss law constraint (\ref{var1}): the latter one  cannot be considered as
 a physical source of the Coulomb potential. 

\medskip At least, it is possible the situation when the transverse (gauge invariant) potential ${\bf A}^D$ is a stationary field:  $\dot{\bf A}^D=0$.
In this case the first item in the  gauge invariant QED Lagrangian density 
(\ref {QEDt1}) becomes zero, the D'alembert equation (\ref{Dalam}) comes to the Poisson equation 
\be \label{Poi}
\Delta  A_k^D({\bf x})= \delta_{ki}^D j_i^D({\bf x}), \ee
permitting the $ A_k^D({\bf x})\sim O(1/r)$ ($k=1,2$) stationary solutions \cite{Landau2} (i.e. that having the look of Coulomb potentials \footnote{In this case the apropriate electromagnetic Green function written down in the momentum representation possesses the asymptotical (infrared) behaviour (cf. \S110  in \cite{BLP})
$$ D_c\to 4\pi/q^2 \quad {\rm as} \quad 
q^2\to 0 \quad {\rm and} \quad \omega=0.$$
}, while the temporal component of a gauge invariant four-potential $A^D$ is also a  Coulomb field that is, indeed, the solution to the Poisson equation $$ A_0^T=\frac{1}{\Delta} j_0^D( {x}),$$ also possessing the $O(1/r)$ behaviour. 
 
As a result, we come to the "pure" electrostatic (if $j_0^D( {x})=j_0^D( {\bf x})$), now given in terms of gauge invariant transverse four-vectors $A^D=(A^T_0, {\bf A}^D)$: 
\be \label{tAD}
p\cdot A^D=0.
\ee
\medskip Indeed, this "pure" electrostatic should be supplemented by a stationary (electrostatic) solution $ A_0({\bf x})$ to the homogeneous Laplace equation \cite{Nguen2}
\be \label{Gauus- Laplace}
\Delta  A_0({\bf x})=0, \ee
that is  the Gauss law constraint (\ref{var1}) resolving in terms of transverse Dirac variables $ A_k^D$: (\ref {**}), (\ref{vak}), in the purely electromagnetic theory (\ref {LEDt}) \cite{Nguen2} (herewith temporal components of these Dirac variables vanish due to their removal (\ref {udalenie}) \cite{David2}).

The homogeneous Laplace equation (\ref{Gauus- Laplace}), as it is well known (see e.g. \S36 in \cite{Landau2}), describes  electrostatic fields $\bf E$ in emptiness:   
\be \label{zakon Kulona}
{\bf E}= \frac{e{\bf R}}{R^3}, \quad A_0({\bf R})\equiv \phi ({\bf R}) =\frac{e}{R},
\ee
obeyed the Coulomb law.

Just these electrostatic fields $\bf E$ and electrostatic Coulomb potentials $\phi ({\bf r})$ \cite{Landau2} supplement the above described picture of electrostatic in terms of transverse Dirac variables $A^D$, (\ref {tAD}).

Thus the solution to the Gauss law constraint (\ref{var1}) can be represented as the sum of the (general) solution (\ref{zakon Kulona}) to the  homogeneous Laplace equation (\ref{Gauus- Laplace}) and the (particular) solution \cite{Pervush2,Azimov} (\ref {var1s}),  (\ref {var1si}) to the inhomogeneous equation (\ref{var1}).

\medskip Resuming the said above, one can speak that  constructing  transverse and physical Dirac variables in QED has the following common  feature with  constructing  transverse and physical fields in usual electrodynamics on-shell of photon \cite{A.I.,Landau2,Rohrlich}. In the both cases one removes  temporal components of  Abelian fields resolving the Gauss law constraint  (\ref{var1}) with the transverse Coulomb gauge. 
But in the Dirac fundamental quantization method  \cite{Dir,Pervush2,Azimov, Nguen2} one utilizes the nonlocal functionals 
(\ref {**}), (\ref{vak}) \cite{Pervush2} of gauge fields. These functionals are gauge invariant and  transverse Dirac variables.

\medskip As we have noticed above, there is a definite ambiguity in specifying Dirac variables (\ref {vak}). They are  specified indeed to within the stationary phase (\ref{faza}), extracting the  subgroup of  stationary gauge  transformations in the general $U(1)$  gauge group inherent in QCD. 

The  stationary phase (\ref{faza}) is fixed    via the additional constraint in the form of the   time integral \rm of the Gauss law constraint~(\ref{gc2}) getting from (\ref {var1}) upon setting $a_0$ in zero by the gauge  transformations (\ref{udalenie}) \cite{LP1,David2}: 
 \be \label{ngauge}  \partial_i A_i^D=0.  \ee  
We shall refer to this equation as to the  constraint-shell or to the \it Coulomb \rm gauge. 

This gauge  restricts initial data to within a phase specifyed by
the equation 
\be 
\label{neodnoznachost}
 \Delta \Phi_0 ({\bf x})=0,
\ee 
that is the spatial part  of the condition
$  \partial _\mu \partial^\mu\Lambda =0$, (\ref{gauge}) \cite{A.I.}  \footnote{Mathematically, Eq. (\ref {neodnoznachost}) also coincides with the Laplace equation (\ref {Gauus- Laplace}) and with the equation \cite{Pervush2} 
$$  \Delta \lambda_L (t,{\bf x})=0  $$ imposed onto the Lagrange multiplier $\lambda_L (t,{\bf x})$ entering the effective 
action  (\ref {efac}).
}.

Nontrivial solutions to this second-order differential equation in partial derivatives we shall call \it the degeneration  of initial data \rm or \it  Gribov copying {\rm \cite{Gribov} } the constraint-shell gauge\rm. 

 Mathematically, solutions to Eq. (\ref {neodnoznachost}) always can be represented \cite{Nguen2}  as 
\be \label{homogen} \frac{c_1}{r}+c_2,\ee
with $c_1$ and $c_2$ being constants. 

\bigskip Finishing our discussion about constructing Dirac variables in constraint-shell four-dimensional QED \cite{Pervush2,Azimov, Nguen2} we should like to make the following important remark.

Constraint-shell QED model (\ref {QEDt1}) \cite{Azimov, Nguen2}  has only  three subtle differences from the initial gauge theory (\ref{QEDa}). \par 
First of them is the origin of the current conservation law. 
 In the initial constrained system (\ref{QEDa}) the current conservation law  
 $$\partial_{0}j_{0}=\partial_{i}j_{i}$$ 
may be derived from the equations for the gauge fields (\ref{var2}) \cite{Landau2} \footnote{As there was demonstrated in \cite{Landau2}, the Maxwell  equations
$$ \frac{\partial F^{ik}}{\partial x^k}=- 4\pi j^i  $$
is mathematically equivalent to
$$ \frac{\partial^2 F^{ik}}{\partial x^i\partial x^k }=- 4\pi \frac{\partial j^i}{\partial x^i}.   $$
But the symmetric  in the indices $i$, $k$ operator $\frac{\partial ^2}{\partial x^i\partial x^k}$, being applied to the antisymmetric Maxwell strength tensor $ F^{ik}$, sets its identically in zero; thus one comes to the continuity  equation (current conservation law) \cite{Landau2}
$$ \frac{\partial j^i} {\partial x^i}=0.   $$
},  whereas the  similar law
 $$\partial_{0}j^D_{0}=\partial_{i}j^D_{i}$$  in 
 EUS (\ref {QEDt1}) follows only from the "Dirac" equation (\ref {Direq}) \cite{Nguen2} for  fermionic  fields (cf. \S 7.5 in \cite{A.I.}) and from the explicit look $e~\bar \psi^D\gamma_\mu\psi^D $ of  gauge invariant fermionic currents, written down in terms of Dirac variables $\bar \psi^D $, $\psi^D $: (\ref {vak}). This difference becomes essential  in quantum theory. \par 
 \medskip In the   case of  constraint-shell four-dimensional QED \cite{Pervush2,Azimov, Nguen2}, involving   EUS (\ref {QEDt1}), generating the "Dirac" equation (\ref {Direq}), we cannot use  the current conservation law when \it  quantum fermions are off-shell\rm:  in particular, in an atom. \par 
 What one may observe in an atom? The bare fermions, or {\it  dressed} ones,  (\ref{vak})? 
 
 Dirac supposed \cite{Dir} that we may observe  only {\it gauge invariant} quantities of the type of  {\it  dressed} fields. \par 
 Really, we may convince ourselves (and this was done above) that the { dressed} fields~(\ref{vak}), as nonlocal functionals of initial gauge fields,  are   invariant with respect to the time dependent gauge transformations   of these initial fields:  (\ref{gauge}),  (\ref{gauge'}),  (\ref{3.9}). \par
 The gauge  invariance with respect to the time dependent gauge transformations  is the second difference of  nonlocal Dirac
 variables~(\ref{vak}) and EUS (\ref {QEDt1}) \cite{Nguen2}  from the constrained system~(\ref{QEDa}) involving ordinary  transformational properties with respect to gauge and Lorentz  transformations (see below).\par
 The gauge constraint $ \partial_{i}A_i=0$ in the gauge-fixing method is associated (see the next subsection) with the relativistic non-covariance. In turn, the observable nonlocal variables~(\ref{vak}) depend on
the time axis in the {\it relativistic covariant} manner. This is the third difference  of the {\it constraint-shell dynamic variables}~(\ref{vak}) in EUS (\ref {QEDt1}) from those in the gauge-fixing method.\par
The gauge-fixing method and its terminology "the Coulomb gauge"  indeed do not reflect these three properties of  Dirac observables in
   constraint-shell QED \cite{Pervush2,Azimov, Nguen2}: the off-shell  non-conservation of the current, gauge invariance with respect to the transformations  (\ref{gauge}), (\ref{gauge'}),  (\ref{3.9}) and relativistic covariance.\par
 In fact,  the term {\it gauge} (for example, the Coulomb gauge
(\ref{ngauge}))  means  a {\it choice of nonlocal  variables}, or more exactly, a {\it a choice of a gauge of physical sources}  associated with these variables.

\subsection{Relativistic covariance of Dirac variables in QED and minimal quantization scheme.}
Dirac variables prove to be manifestly relativistically covariant.  
  Relativistic properties of Dirac variables in gauge theories were investigated in the papers \cite{Heisenberg} (with the reference to the unpublished note by von Neumann), and  then  this job was continued by I. V. Polubarinov in his review
 \cite{Polubarinov}. These investigations displayed that there exist such relativistic transformations of Dirac variables that maintain transverse gauges of  fields. 
 
 Let us demonstrate this now with the example of  constraint-shell QED, us discusssed in the previous subsection. If one makes therein the usual relativistic transformations of  initial
 fields $A_i, A_0, \psi  $ with the parameter $\epsilon_i$ \cite{Pervush2}:
  \begin{eqnarray}
\label{ultf}  
 \delta_{L}^{0} A_{k}  &=& \epsilon_{i} ( x_{i}^{\prime}  \partial_{0^{\prime}} - x_{0}^{\prime} \partial_{i^{\prime}} )  A_{k}(x^{\prime}) + \epsilon_{k} A_0,  \nonumber \\ \\   \delta_{L}^{0} \psi &=& \epsilon_{i} ( x_{i}^{\prime}  \partial_{0^{\prime}} - x_{0}^{\prime} \partial_{i^{\prime}} )  \psi ( x^{\prime} ) + \frac{1}{4} \epsilon_{k} [\gamma_{i},
 \gamma_{j} ] \psi (x^{\prime} ),  \nonumber 
 \end{eqnarray}
 then the physical Dirac variables~(\ref{vak})  suffer the \it Heisenberg-Pauli  transformations \rm \cite{Heisenberg} 
  \begin{eqnarray}
\label{ltf} 
 A_{k}^{D} [ A_{i} &+& \delta_{L}^{0} A ] - A_{k}^{D} [ A ]  =  \delta_{L}^{0} A_{k}^{D} + \partial_{k} \Lambda,  \end{eqnarray} 
 \begin{eqnarray}
\label{ltf1}
 \psi^{D} [ A &+& \delta_{L}^{0} A , \psi + \delta_{L}^{0} \psi ] -  \psi^{D} [ A, \psi ] = \delta_{L}^{0} \psi^{D} + i e \Lambda  (x^{\prime}) \psi^{D},  \end{eqnarray}  
 with 
 \begin{eqnarray} \label{lqed}
 \Lambda [A^D,j^D_0] = \epsilon_{k} \frac{1}{ \Delta } ( \partial_{0}  A_{k}^{D} + \partial_{k} \frac{1}{\Delta} j_{0}^{D}).
 \end{eqnarray}  
These transformations were interpreted as the  transition from the Coulomb gauge with respect to the time axis in  the fixed {\it rest } reference frame \(\eta_{\mu}^0=(1,0,0,0)\) to the Coulomb gauge  with respect to the time axis in a {\it moving } reference frame (see Fig.1) \cite{Pervush2}
 \be \label{Lor} 
  \eta'_{\mu}=\eta_{\mu}^0 + \delta_L {\eta_{\mu}}^0 \;=\;  {(L\eta^0)}_{\mu}.\
\ee
 \unitlength=1mm \special{em:linewidth 0.4pt} \linethickness{0.4pt} \begin{picture}(110.00,130.00) \put(17.00,100.00){\vector(0,1){30.00}}
\put(50.00,100.00){\vector(0,1){30.00}}
\put(80.00,100.00){\vector(1,3){10.00}} \put(114.00,100.00){\vector(1,3){10.00}} \put(63.00,116.00){\line(1,0){12.00}} \put(63.00,114.00){\line(1,0){12.00}}
\put(71.00,119.00){\line(5,-4){5.00}} \put(76.00,115.00){\line(-5,-4){5.00}}
\put(32.00,114.00){\makebox(0,0)[cc]{${\eta^0}=(1,0,0,0)$}} \put(99.00,114.00){\makebox(0,0)[cc]{${\eta}^{\prime}={\eta^0 }+\delta^0_L{\eta^0}$}} \put(58.00,121.00){\makebox(0,0)[cc]{$A_0^T=0$}} \put(138.00,121.00){\makebox(0,0)[cc]{$A_0^{T^{\prime}}=(\eta^{\prime} \cdot A) =0$}} \put(64.00,105.00){\makebox(0,0)[cc]{$\eta=(1,0,0,0)$}} \put(128.00,105.00){\makebox(0,0)[cc]{$\eta^{\prime}= \eta
+\delta^0_L \eta $}} \put(20.00,90.00){\makebox(0,0)[cc]{ \tt Lorentz }} \put(17.50,85.00){\makebox(0,0)[cc]{\tt Frame}} \put(52.00,90.00){\makebox(0,0)[cc]{\sl Gauge}} \put(82.00,95.00){\makebox(0,0)[cc]{\underline {New} }} \put(85.00,90.00){\makebox(0,0)[cc]{\tt Lorentz }} \put(81.95,85.00){\makebox(0,0)[cc]{\tt Frame}} \put(116.00,95.00){\makebox(0,0)[cc]{\underline {New} }}
 \put(117.00,90.00){\makebox(0,0)[cc]{\sl Gauge}}  \end{picture}  \vspace{-8.4cm}  \begin{center}
{\bf Figure 1.}  \end{center}   
These transformations correspond to the "change of variables"  
\begin{eqnarray} 
\label{cgauge}
 \psi^D(\eta), \hat A^D(\eta) \rightarrow \psi^D(\eta^{\prime}),
 \hat A^D(\eta^{\prime}),   \end{eqnarray} 
 so that these variables become transverse with respect to the new time axis $\eta'$  (or, from the point of view of the "gauge-fixing" method of reduction, with respect to  the transformations~(\ref{ultf}),~(\ref{ltf}),  (\ref{ltf1})) corresponding  to the  "change of a gauge": 
$$ \partial_i {\hat A}^{D\prime}_i=0; \quad i=1,2.; \quad A_0^T=0.
$$

\medskip In the recent paper \cite{math} it was argued that it is possible to extract the so-called  'small' subgroup ${\cal G}_0$ of Poincare invariant gauge transformations from the  'large' group ${\cal G}$ of all the gauge transformations in constraint-shell QED acting in the space of Dirac variables $ A^D$. Such 'small' subgroup ${\cal G}_0$ can be set by the condition 
 \be \label{small1}
 \delta_{L}^{0} A_{k}^{D}=-\partial_{k} \Lambda,
 \ee
 following directly from Eq. (\ref{ltf}) for the apropriate Dirac variables $A^{D}$.

\medskip
As an important result,  one gets (in the rest reference frame $\eta^0$)  the relativistic covariant  separation of the interaction with the Coulomb potential \rm  (instantaneous with respect to the time axis $\eta^0_\mu$)  and   retardations \footnote{One can neglect  retarded interactions if \cite{Dub} 
$$
  |x^0_{\rm (out)}-x^0_{\rm (in)}|\gg E^{-1}_{I,\rm min}, \quad
  ~V_0^{1/3}\gg  E^{-1}_{I,\rm min}
  .$$
  This condition means that all stationary
 solutions
 with  zero energy $E_{I,\rm min} \to 0$ cannot
 be considered as  perturbatios.} (say, (\ref{retard})).
 
 The Coulomb interaction  herewith takes the manifest covariant look
\begin{eqnarray} 
\label{rc1} 
 { W}_{C} = \int d^4 x d^4 y \frac{1}{2} j_{\eta}^{D}(x) V_C(z^{\perp})  j_{\eta}^{D}(y) \delta(\eta \cdot z) .  
\end{eqnarray} 
 Here  
 \begin{eqnarray} 
\label{ccar1} 
 j_{\eta}^{D} = e \bar {\psi}^D \rlap/\eta \psi^D, \quad  z_{\mu}^{\perp} = z_{\mu} - \eta_{\mu}^0(z \cdot \eta) ,\quad  z_\mu = (x - y)_\mu,  \quad  \rlap/\eta \equiv \eta_\mu \gamma ^\mu,  
\end{eqnarray}  
\begin{eqnarray} 
\label{cf}
 V_C(r) = - \frac{1}{4\pi r},\quad r = \vert {\bf z} \vert.  
\end{eqnarray} 

Knowing  an instantaneous bound state (\ref{rc1}) in constraint-shell QED, one can give the definition of relativistic (Lorentz) invariant states. 

As it was noted in  Ref.  \cite{Dub},  the relativistic invariance
 means that {\it a complete set of states
 $\{|\Phi_I>\}_{n^{\rm cf}}$
 obtained by all Lorentz transformations
 of  a state $|\Phi>_0$
 in a  rest frame  $\eta^{0}_\mu=(1,0,0,0)$
 coincides with a complete set of states $\{|\Phi_I>\}_\eta$
 obtained by all Lorentz transformations of this state $|\Phi>_0$
 in another   reference frame} $\eta_\mu=
 (\frac{1}{\sqrt{1-\vec v^2}},\frac{\vec v}{\sqrt{1-\vec v^2}})$.
 
  Issuing from this assertion, it is enough to find all the Lorentz transformations of the rest frame $\eta^{0}_\mu$,  i.e. all the Lorentz transformations
 of the instantaneous interaction (\ref{rc1}) obtained
in this rest frame. 

\medskip
The finite Lorentz transformations from the time axis $\eta^{(1)}$ to the  time axis $\eta^{(2)}$ were constructed in the paper \cite{Polubarinov} using the gauge transformations 
 \be \label{zamena mashtaba}  ieA^{(2)}=U_{(2,1)}[ieA^{(1)}+\partial ]U_{(2,1)}^{-1},\quad  \psi^{(2)}=U_{(2,1)} \psi^{(1)},  \ee   where 
 $$U_{2,1}=v_{(2)}v_{(1)}^{-1}$$ 
and $v_{(2)},v_{(1)}$ are the
 Dirac gauge factors~(\ref{va1})  associated with the time axes $\eta^{(2)}$  and $\eta^{(1)}$, respectively.\par 
Returning to the  initial QED action (\ref{QEDa}), again note that it is not compatible with
 quantum principles, as it contains the zero canonical momentum  by temporal component of the electromagnetic field. 

 The Dirac formulation \cite{Dir} of EUS (\ref{QEDt1}) in four-dimensional QED \cite{Pervush2, Azimov,Nguen2} keeps the quantum  principles by  values that exclude the unphysical components. 

One quantizes then  EUS (\ref{QEDt1})  involving gauge invariant physical Dirac variables~(\ref{vak}).

The apropriate commutation relations \cite{Pervush2,Nguen2} 
 \begin{eqnarray*} 
 i \bigl[ \partial_{0} A^{D}_i ( {\bf x}, t ), ~~ A^{D}_j (  {\bf y}, t ) \bigr]  &=& ( \delta_{ij} -
 \partial_{i}  \frac{1}{ \Delta } \partial_{j} )   \delta^3 ( {\bf x} - {\bf y} )\equiv \delta_{ij}^D\delta^3 ( {\bf x} - {\bf y}),   \\  \nonumber  \\  \bigl\{  {\hat \psi}^{D +} ({\bf x},t ) , \hat {\psi}^{D} ( {\bf  y},t ) \bigr\} &=& \delta^3 ( {\bf x} - {\bf y} )  \end{eqnarray*}
  lead to the generating functional for Green's function of the  obtained unconstrained system in the form of the   Feynman path
 integral in the fixed  reference frame $\eta$ \cite{Pervush2, Feynman1}:
 \be \label{fi}
 Z_{\eta}^{*}[ s^*, {\bar {s}}^*, J^* ]\;=\;\int \prod_{j} DA^*_j
 D\psi^*D{\bar \psi}^*
 e^{iW^{*}[A^*,\psi^*, {\bar \psi}^*] + i S^* },
 \ee
 including the external sources term:
  \be \label{si}
 S^*\;=\;\int d^4x \left({\bar s}^* \psi^* + {\bar \psi}^* s^*
 +J^*_i A^{*i} \right)
 \ee
(here $A^*$ and $\psi^*= v({\bf x})\psi$ are Dirac variables for gauge
and fermionic fields, respectively; $J^*$ is the source of gauge
 fields,
${\bar s}^*$ and $s^*$ are the sources of  fermionic fields $\bar \psi^*$ and $\psi^*$, respectively; $W^*$ is
the \it constraint-shell \rm action of the considered theory \footnote{Besides the above commutation relations, also it is worth to cite here the following ones \cite{Nguen2}, taking place in  constraint-shell QED \cite{Pervush2, Azimov,Nguen2}. So, 
$$ i \bigl[ \partial_{0} A^{D}_i ( {\bf x}, t ), ~~ A^{D}_j (  {\bf y}, t ) \bigr] =\delta_{ij}^D\delta^3 ( {\bf x} - {\bf y}) \Longleftrightarrow  i \bigl[F_{0i}^D ( {\bf x}, t ), ~~ A^{D}_j (  {\bf y}, t ) \bigr] =\delta_{ij}^D\delta^3 ( {\bf x} - {\bf y}),  $$ 
with 
$$ F_{0i}^D= \dot A_i^D- \partial_i A_0^T.  $$
The latter commutation relation follows from manifest commutations of bosonic fields $ A_i^D $ ($i=1,2$) and fermionic charges $j_0^D$ due to
$$ \bigl[ \psi^D ( {\bf x}, t ), ~~ A^{D}_i ( {\bf y}, t ) \bigr] =0.  $$
On the other hand,
$$  \bigl[A_0^T ( {\bf x}, t ),~~ \psi^D ( {\bf x}, t ) \bigr] \sim - \frac{1}{4\pi \vert {\bf x}-{\bf y}\vert } \psi^D ( {\bf y})    $$
in the equal time instant $x_0=y_0=t_0$.
}.

By  constructing  the  unconstrained system, this generating  functional is manifestly {\it gauge invariant} and {\it relativistic covariant} (due to the  theory (\ref{ltf}), (\ref{ltf1})).

\medskip Relativistic transformation properties of   quantum fields would repeat the ones of the Dirac  variables~(\ref{vak}) as nonlocal functionals
 of the initial fields. 

 As it was shown in the papers \cite{Pervush2,Bogolubskaja, Polubarinov,Schwinger2,Zumino,  Schwinger1,  Ozaki,Smorodinskij}, a quantum theory (say, 
four-dimensional QED)  involving the gauge invariant \it  Belinfante energy-momentum tensor \rm 
\begin{eqnarray} \label{beli}
 T_{ \mu \nu}  & = & F_{\mu \lambda }F_{\nu}^{\lambda} + {\bar  {\psi}} \gamma_{\mu} [ i\partial_{\nu} + e A_{\nu} ] \psi
  -  g_{\mu \nu} L + \nonumber \\  & + & { i \over 4 }\partial_{\lambda} [ \bar {\psi}  \Gamma_{\mu\nu}^{\lambda} \psi ],
 \end{eqnarray}   \begin{eqnarray*}  \Gamma_{\mu\nu}^{\lambda} = {1\over  2}[\gamma^{\lambda}\gamma_{\mu}]\gamma_{\nu} - g_{\mu\nu }
 \gamma^{\lambda} - g_{\nu}^{\lambda}\gamma_{\mu},  
\end{eqnarray*} 
on the surface of the Gauss law constraint (\ref{var1}), i.e.  being rewritten in terms 
of the Dirac variables \cite{Bogolubskaja}:  
\begin{eqnarray} \label{beli1}
 T_{ \mu \nu} \left [A_i, A_0=   \left ( \frac{1}{\Delta} \partial_i \partial_0 A ^i +j_0\right ) \right] =  T_{ \mu \nu} [A^D[A_i], \psi ^D[A,\psi],\bar \psi ^D[A,\psi]],  
\end{eqnarray} 
completely reproduces the symmetry properties of the "classical"
 theory  (\ref{ultf})-   (\ref{lqed}):
  \bea \label{ql}
 i\epsilon_k[M_{0k},\psi^D]&=&\delta^0_L\psi^D+ie\Lambda[A^D,j^D_0]\psi^D; \nonumber \\
i\epsilon_k[M_{0k},A^D_\mu (x)] &=& \delta^0_L A^D_\mu (x)+ \partial_\mu \Lambda;  \nonumber \\   M_{0k}&=&\int d^3x[x_kT_{00}-tT_{0k}].  \eea
For us there will be very useful to cite now the explicit look of $T_{00}$ and $ T_{0k}$, entering the expression (\ref{ql})  for the Lorentz boost $ M_{0k}$
 in  constraint-shell QED \cite{Pervush2, Azimov,Nguen2}.
 
 So, $T_{00}$ is determined by the explicit look of the constraint-shell gauge invariant Lagrangian density (\ref{QEDt1}). Substituting the latter one in the expression (\ref{beli}) for the Belinfante energy-momentum tensor, we get 
\cite{Nguen2}
\be
\label{beliQ}
T_{00}=\frac{1}{2} (F_{0i}^D)^2 +\frac{1}{4} F_{ij}^2 +\bar \psi ^D (i\gamma_i \partial_i -m ) \psi ^D. 
\ee
Furthermore \cite{Nguen2},
\be
\label{Tok}
T_{0k}= F_{0i}^D F_{ki} + \psi^{+D} \gamma_0 \partial_k \psi ^{D}  + \frac {i}{4} \partial_i (\psi ^{+D} \gamma_0 [\gamma_i,~~\gamma_k]) \psi ^{D}.
\ee
Herewith there is  implemented the commutation relation \cite{Nguen2}
\be
\label{priznak}
i \bigl [ T_{00}(x), ~~ T_{00}(y) \bigr] =- (T_{00}(x)+ T_{00}(y)) \partial_k \delta ^3 ~( {\bf x}-{\bf y}).
\ee 
Generally speaking, as there was explained in the paper \cite{Schwinger1}, implementing the  commutation relation of the 
(\ref{priznak}) type in a gauge theory is the necessary condition for this theory to be manifestly relativistic (Lorentz) invariant \footnote{ In this case the commutation relation (\ref{priznak}) is equivalent to the appearance of the {\it Schwinger surface terms } 
$$\sim S^{ab}_{ij}({\bf x})\int \frac{\partial }{\partial y_i}\delta^3({\bf x}-{\bf y})d^3y=0$$
in  action functionals of gauge theories. 
This, in turn, implies their manifest Lorentz invariance. 
}.

However, in four-dimensional constraint-shell QED \cite{Pervush2, Azimov,Nguen2}  the commutation relation (\ref{priznak}) follows directly
from explicit solving the Gauss law constraint (\ref {var1}) in terms of the Dirac variables (\ref {vak}). 

Thus the commutation relation (\ref{priznak}) acquires in constraint-shell QED \cite{Pervush2, Azimov,Nguen2} the rather  another sense than in the Schwinger model \cite{Schwinger1} for describing gauge theories.

Moreover, there may be directly checked \cite{Nguen2} 
that if the operators  
\be
\label{Gammi}
H=\int d^3x~ T_{00},
\ee
\be
\label{poka}
P_k = \int d^3x~ T_{0k} \ee and also the momentum operators $M_{0k}$, $M_{ij}$ satisfy the ordinary algebra \cite{Nguen2} of the Poincare group generators in the physical sector, 
(\ref {vak}), of gauge fields in constraint-shell QED \cite{Pervush2, Azimov,Nguen2} (we omit these  commutation relations, since they are, perhaps, well-known to our readers), then  the commutation relation (\ref{priznak}) is automatically fulfilled.

\medskip It will be also useful to  write down explicitly the commutation relations between the momentum operator $P$, given in (\ref {poka}), and Dirac variables $A_\mu^D$ and $\psi^D$, given in (\ref {vak}).

These are \cite{Nguen2}
\be \label{4169}
i[P_\mu, A_\nu ^D]= \partial _\mu A_\nu ^D, \quad  i[P_\mu, \psi^D]= \partial _\mu \psi^D.
\ee
Finally \cite{Nguen2}, the Poincare algebra of the operators  $ H$, $P_k$, $M_{0k}$ and $M_{ij}$, supplemented by the commutation relation (\ref{priznak}), just results the  Heisenberg-Pauli  transformations \cite{Pervush2} (\ref {ltf})-  (\ref{lqed}), accompanied, in turn, by the Lorentz transformation (\ref {Lor}) of the chosen (rest) reference frame.

Thus  the commutation relation (\ref{priznak}) in four-dimensional QED \cite{Pervush2, Azimov,Nguen2} (unlike the analogous one in the  Schwinger model \cite{Schwinger1} for describing gauge theories) cannot serve as the sufficient condition for the {\it relativistic invariance} of the given theory. Rather the opposite effect takes place, {\it the manifest relativistic covariance}  of four-dimensional QED  \cite{Pervush2, Azimov,Nguen2}.

\medskip We should like make the following concluding remark \cite{Nguen2} concerning  the commutation relations in four-dimensional constraint-shell QED \cite{Pervush2, Azimov,Nguen2}.

At going over from a classical to quantum  theory, apart from defining simultaneous commutation relations, one must also eliminate pseudophysical quantities like a zero energy, a zero charge, etc.

To perform this elimination, one uses, as a rule, a normal product \cite {Bogolubov-Shirkov}
for dynamical variables, depending quadratically on operators with identical arguments.

However \cite {Schwinger1,Xriplovich}, utilizing N-product becomes an unnecessary although harmless thing in the spinor electrodynamics (generally speaking, it contradicts the gauge invariance in scalar electrodynamics and in the YM theory).

To achieve the manifest  gauge invariance of a theory, it is sufficient to symmetrize the apropriate Hamiltonian in Bose operators and  antisymmetrize it by Fermi operators, i.e. to use the Weyl quantization.
	 
It is well known that this type of presentation of current densities $j_\mu$ of spinor and scalar particles implies that particles and antiparticles enter $j_\mu$ symmetrically, and the vacuum expectation value of $j_\mu$ is equal to zero.
It is just the recipe we shall follow in the present study.

For instance, in  the case \cite{Pervush2, Azimov,Nguen2} of four-dimensional constraint-shell QED (that is the spinor electrodynamics), the symmetrization procedure of the operators $A_i^D$ ($i=1,2$), $F_{0i}(A^D)$ (given in (\ref {QEDt}) \cite{Nguen2}) and $\psi^D$ does not influence the apropriate (constraint-shell) Hamiltonian $H$ (\ref {Gammi}), momentum tensor $M_{i, j}$ and boost tensor $M_{0i}$ (given in (\ref {ql}) \cite{Pervush2}).

\bigskip
  The Lorentz transformation (\ref {Lor}) of the chosen (rest) reference frame, at the level of the operator quantization, including   the apropriate Feynman path integral, means the change of the  time axis:  
\be
\label{rfi}  
Z_{L\eta}^{*}[ s^*, {\bar {s}}^*,J^* ]\;=\; Z_{\eta}^{*}[  Ls^*, L{\bar {s}}^*,LJ^* ].
 \ee  
This scheme of quantization, called \it the minimal quantization scheme \rm \cite{ Azimov,Nguen2,Bogolubskaja,Nguen1, Nevena, Werner}, \footnote{As the additional claim to the minimal quantization scheme, one would dioganalize the Belinfante Hamiltonian $T_{00}$. Then, on the operator level, the Dirac variables (\ref{vak}) coincides with the quantum ones.} explicitly depends on a choice of  the time axis.

 If one choose a definite reference frame with  the initial time axis, any Lorentz transformation  turns this time axis in a relativistic covariant manner. 
This   implies that {\it constraint dynamics}  is manifestly {\it relativistic covariant}. 

 Another problem is to find conditions at which measurable physical  quantities and results of theoretical calculations
 do not depend on the time axis (identified with a physical device \cite{LP2}).
 
 This independence exists  only for scattering amplitudes   of particles on-shell \rm \cite{fund,Fadd1}. In this case one  may speak about the {\it relativistic invariance} of  scattering amplitudes squared as functionals of local degrees of freedom.

 But it is well known that Green functions (in particular,  one-particle Green functions) and instantaneous bound states depend on  the choice of the time axis. 

In general case,  measurable  quantities in electrodynamics  depend on the time axis and other parameters of a physical device,  including its size and energy resolution \cite{Pervush3}. \par
If a nonlocal process depends on the time axis, one should  to establish a principle how to  choose  this time axis. \par  \medskip
  In particular, this choice and the nonlocal relativistic
 transformations (\ref{ql})  remove all the  infrared divergences from
 the one-particle Green function (for instance, in four-dimensional constraint-shell QED \cite{Pervush2, Azimov,Nguen2}  written in terms of the radiation  variables in the rest reference frame of an electron $p_\mu=(p_0,0)$ for the time axis $l_\mu^0=(1,0,0,0)$ \cite{Pervush2,Nguen2,Nevena, Yura2}:
  \be \label{green1}  
i (2\pi)^4 \delta^4(p-q) G(p-q)=\int\limits_{ }^{ } d^4d^4y  \exp(ipx-iqy)<0|T\bar \psi^D(x)\psi^D(y)|0>,  \ee  
$$  G(p)=G_0(p)+ G_0(p) \Sigma (p) G_0(p) + O(\alpha^4), \quad  G_0(p)=[\not p-m]^{-1}  $$
 $$  \Sigma (p)=\frac{\alpha}{8\pi^3 i}\int\limits_{ }^{ }  \frac{d^4q }{q^2+i\epsilon}\left[  \left(\delta_{ij}-q_i  \frac{1}{\vec q^2} q_j  \right)
 \gamma_i G_0(p-q)\gamma_j    + \gamma_0  G_0(p-q)\gamma_0\frac{1}{\vec q^2}\right]=  \frac{\alpha}{4\pi}\Pi(p),  $$
 where $\Pi(p)$ is 
 $$
  m(3D+4)-D (\not p-m)+\frac{1}{2}  (\not p-m)^2\left[\frac{(\not p+m)}{p^2} \left(\ln\frac{m^2-p^2}{m^2} \right)  \left(1+\frac{\not p(\not p-m)}{2p^2}\right)-  \frac{\not p}{2p^2}\right]
 $$
 and $D$ is the ultraviolet dimensional-regularization parameter \cite{Nguen2}:
$$ D=1/\epsilon- \gamma_\epsilon +\ln 4\pi.$$ Herewith there may be adopted, as it is customary,  $\epsilon =4-d$, where $d$ is the dimension of the space-time. 

We  also recommend our readers \S\S~ 9.5-  9.7 in \cite{Ryder} where the dimensional regularization of (one-loop) QED was detailed described. At computing the mass operator  $\Sigma (p) $ \cite{Pervush2, Nguen2}, (\ref{green1}), in   four-dimensional constraint-shell QED \cite{Pervush2, Azimov,Nguen2} the similar methods were utilized.

As it is customary in QED, two items in the mass operator  $\Sigma (p)$ in (\ref{green1}) describe, respectively, contributions from transverse Dirac fields $A_i^D$ ($i=1,2$) and from temporal Dirac components $A^T$ to the apropriate Green function (\ref{green1}).

Indeed, there may be demonstrated the manifest relativistic invariance of the mass operator  $\Sigma (p)$ in constraint-shell QED \cite{Pervush2, Azimov,Nguen2} with respect to the Lorentz (Heisenberg-Pauli)) transformations (\ref {ltf})-  (\ref{lqed}) of gauge fields $A^D_i$ ($i=1,2$) and fermionic ones $\psi^D$, $\bar\psi^D$ \cite{Heisenberg,Pervush2, Nguen2,Polubarinov}:
\be \label{Ltot}
\delta_L^{\rm tot} \Sigma (p)=( \delta_L^{0}+\delta_\Lambda) \Sigma (p)=0.
\ee
To prove the statement about the relativistic invariance of  the mass operator  $\Sigma (p)$ in constraint-shell QED \cite{Pervush2, Azimov,Nguen2} note firstly that the fermionic Green function $ G_0(p-q)\equiv \tilde G_0$ entering the mass operator  $\Sigma (p)$, (\ref{green1}), is just manifestly  invariant with respect to the Lorentz  transformations $\delta_L^{0}$ since $\not p$ and $\not q$ possess such property. 
This statement can be written down as $\delta_L^{0}G_0(p)=0$ \cite{Nguen2}. 

Then we rewrite $\Sigma (p)$ as the sum
\be \label{sigmap}
\Sigma (p) = \Sigma _F(p) + \Delta \Sigma (p), \ee
where $\Sigma _F(p) $ is the mass operator  (electron self-energy) in the Feynman gauge (see \S 76 in \cite{BLP})
\be \label{Feynman gauge}
\Sigma _F(p)= - \frac{ie^2}{32\pi^4} \int \frac{d^4q}{q^2+i\epsilon} \gamma_\mu  \tilde G_0 \gamma_\mu,  
\ee
invariant with respect to the Lorentz  transformations $\delta_L^{0}$ ($\delta_L \Sigma _F(p)=0$), while the noninvariant addition $\Delta \Sigma (p)$ has the look
\be \label{dsig}
\Delta \Sigma (p) =  \frac{ie^2}{32\pi^4} \int \frac{d^4q}{q^2 {\bf q}^2} [\not q \tilde G_0\not q+ \underline q \tilde G_0\not q +\not q \tilde G_0 \underline q]
\ee
$$ \underline q = {\vec \gamma }\cdot {\bf q}$$
(${\vec \gamma }\equiv (\gamma_1, \gamma_2, \gamma_3 )$). 
The response  of $\Delta \Sigma (p)$ to  Lorentz  transformations (\ref {ultf}) may be found by rotations
\be \label{respd}
\delta_L^{0} p_0=\epsilon_k  p_k,  \quad \delta_L^{0} p_k=\epsilon_k p_0, \quad   \delta_L^{0} \gamma_0 = \epsilon_k \gamma_k, \quad \delta_L^{0} \gamma_k = \epsilon_k \gamma_0,
\ee
with $\epsilon_i $ being an infinitesimal Lorentz parameter involving Lorentz boosts 
$$x'_k=  x_k+ \epsilon_ k t, \quad t'= t+ \epsilon_ k x_k, \quad \vert \epsilon_ k \vert \ll 1. $$
As a result, we can get the Lorentz  transformations for $ \Sigma (p)$ given in the integral representation \cite{Nguen2}:
\bea \label{dsigL} \delta_L^{0} \Delta \Sigma (p)&=& \epsilon_ k \frac{ie^2}{32\pi^4} \int \frac{d^4q}{q^2 {\bf q}^2} [-\frac{2q_0 q_k } {{\bf q}^2} \not q \tilde G_0\not q \nonumber \\ & + & (q_k\gamma_0+\gamma_k q_0- \underline q \frac{2q_0 q_k } {{\bf q}^2}) \tilde G_0\not q+ \not q \tilde G_0 (q_k\gamma_0+\gamma_k q_0- \underline q \frac{2q_0 q_k } {{\bf q}^2})] \nonumber  \\
&=& \epsilon_ k \frac{ie^2}{32\pi^4} \int \frac{d^4q}{q^2 {\bf q}^2} [B_k \tilde G_0\not q +\not q \tilde G_0 B_k],
\eea
with
\be \label{Bk} 
B_k= q_k\gamma_0+\gamma_k q_0- \frac{2q_0 q_k } {{\bf q}^2} \underline q -\frac{q_0 q_k } {{\bf q}^2} \not q.
\ee
However, the total  Lorentz  transformations for the Green function (\ref{green1})
also contain the gauge transformations $\delta_\Lambda$, 
(\ref {lqed}):
\bea \label{dsigLa}
&~ & \delta_\Lambda [(2\pi)^4\delta^4(p-q)iG(p)]\nonumber \\ &=& i e\epsilon_k  \int d^4 x d^4 y e^{ipx- iqy}[<0\vert T(\psi^D(x) \bar \psi^D(y) \Lambda _k (y)) \vert 0> \nonumber \\&-& <0 \vert T(\Lambda _k (x) \psi^D(x) \bar \psi^D(y)) \vert 0> ].
\eea
Using the explicit look (\ref {lqed}) for $\Lambda_k(x)= \Lambda^T_k(x)+ \Lambda^c_k(x)$ with
\be \label {tc}
\Lambda^T_k({\bf x}, t)= - \frac{1}{4\pi}\int d^3 y  \frac{\dot A_k^D({\bf y}, t)}{ \vert {\bf x}-{\bf y}\vert },   \quad \Lambda^T_k({\bf x}, t)= - \frac{1}{4\pi}\int d^3 y  \frac{\partial_k A_0^T ({\bf y}, t)}{ \vert {\bf x}-{\bf y}\vert },
\ee
we get 
\be \label{dsigLa1}
\delta_\Lambda \Sigma (p)=- \epsilon_k \frac{ie^2}{32\pi^4}   \int  \frac{d^4q}{q^2 {\bf q}^2} [B_k \tilde G_0(\not p-m)+ (\not p-m) \tilde G_0 B_k],
\ee
\be \label{Bk1} 
B_k = (\delta_{ki} -q_k \frac{1}{{\bf q}^2} q_i)\gamma_i q_0 -\frac{q_k q_0^2\gamma_0}{{\bf q}^2}+  \gamma_0 q_k.
\ee
It is easy to check that $B_k$ given by Eq. (\ref{Bk1}) coincide with those given by Eq. (\ref{Bk}).

As
\be \label{4178}
(\not p-m)G_0(p-q)= 1+ \not q  G_0(p-q)
\ee
and 
\be \label{4179}
\int \frac{d^4q}{q^2 {\bf q}^2} B_k=0,
\ee
we get the final expression
\bea \label{finale}
&~& \delta_\Lambda [(2\pi)^4\delta^4(p-q) iG(p)] \nonumber \\ &=& -\epsilon_k \int \frac{ie^2}{32\pi^4} \frac{d^4q}{q^2 {\bf q}^2} [B_k \tilde G_0\not q + \not q \tilde G_0 B_k]=- \delta_L^0 \Sigma (p). \eea
The total response of $\Sigma (p)$ to the Lorentz transformations (\ref{ltf})- (\ref{lqed}) is  thus equal to zero:
\be \label{total response}
\delta_L^{\rm tot} \Sigma (p)= (\delta_L^0 +\delta_\Lambda) \Sigma (p)=0. \ee
Thus it is sufficient to calculate the electronic Green function in constraint-shell QED \cite{Pervush2, Azimov,Nguen2} in the rest reference frame of the observable electron, as it was done in (\ref {green1}).

\medskip Since \cite{Nguen2}
\be \label{4182}
\partial_\mu ^l  \partial_\mu - l_\mu (\partial l) \Longleftrightarrow  A^D= A^D- l (A^D \cdot l),
\ee
for the given time-like vector $A^D=(A_0^T A_i^D)$ ($i=1,2$) at a Lorentz transformation (\ref {Lor}),
$$ l'_\mu= l_\mu^0+ \delta_L^0 l_\mu^0,$$ of the rest reference frame $l_\mu^0$,
then apropriate passing
$$p_\mu=(p_0, 0)\to p'_{\mu}=(p'_0,{\bf p}'\not= 0)$$  
in the momentum space, associated with  the local Lorentz transformations 
(\ref {dsigLa}), involves the change of gauge for the Dirac variables $ A^D $ (cf.  Fig. 1):
\be \label{ism. kalibrovki}
q_i A_i ^D (q)= 0 \Longrightarrow  [q -l(ql)] A^D=0,
\ee
where the moving reference frame $ l'$ may be expressed through the momentum $p'_{\mu}$ as
\be \label{4184}
l'_\mu = p'_\mu / \sqrt{p'^2}.
\ee
\medskip It is enough transparent that the local Lorentz transformations 
(\ref {dsigLa}) for the electronic Green function involve new 
Feynman diagrams, referred to as {\it spurious} ones (e.g. in \cite{Pervush2}). 

 As a result,  in another reference frame $ l'$  we get  the same relativistic covariant expressions depending on the new  momentum $p'$ \cite{Nguen2, Yura2}.\par 
\medskip It is  also  worth to note that the electronic self-energy $\Sigma(p)$, given in (\ref{green1})  \cite{Pervush2, Nguen2}, has no infrared divergences and allows the renormalization with subtracting  at physical values of the momentum, $\not p=m$ \footnote{It can be proven that \cite{Nguen2002}
$$\Sigma (\not p =m) =\delta m =\frac {m\alpha}{4\pi} (3D+4)$$
in this case. 
}.

The next important property of the electronic Green function $G(p)$, (\ref{green1}), written down in terms of Dirac variables $A^D$, is  that the probability to find an  electron with the mass $m$ specifying by the formula \cite{Nguen2}
\be \label{prob}
R(p)= \lim_{\not p\to m} (\not p -m) G_R(p)= \vert \psi \vert ^2
\ee
is equal to unity ($\vert \psi \vert ^2
=1$).

It is the consequence of the relativistic invariant look $({\not p}-m)^{-1}$ of the electronic Green function $G(\not p)$ in (\ref{green1}).

The result (\ref{prob}) represents a solution to the renormalization problem on the mass-shell for transverse variables $A^D$. 

\medskip A mistake of the popular papers \cite {Hagen, Bjorken} was not only  ignoring correct transformation properties (\ref {ltf})-  (\ref{lqed}) of Dirac variables $A^D$ in constructing $\Sigma(p)$ in four-dimensional constraint-shell QED \cite{Pervush2, Azimov,Nguen2} but also in a unphysical choice of the time  axis, that, in turn, incorrectly fixes   temporal components of gauge fields, i.e. Coulomb (electrostatic) fields. 

For instance, in Eq. (\ref{green1}),  when $p_\mu=(p_0,{\bf p}\neq 0)$, the vector $l_\mu^0=(1,0,0,0)$ may be chosen so that an electron has the velocity different from that of its Coulomb  field.
As a result, there arise definite difficulties with the manifest Lorentz invariance and infrared divergences. 
On the other hand, the correct transition to the rest reference frame $p_\mu=(p_0,{\bf p}= 0)$ in (\ref{green1}) doesn't remove these difficulties as one simultaneously rotate the initial rest reference frame $l_\mu^0=(1,0,0,0)$,  thus leaving   velocities of an electron and its proper   field to be  different from those in the Coulomb case. 

So, the choice of $ l_\mu^0$ must be specified by  physical formulating the problem; in this  case $ l_\mu^0$ is, indeed, the unit vector along the momentum, $   p_\mu\sim l_\mu^0$. 

\medskip Comparing  the formulas (\ref{green1})  and (\ref{3.17}) \cite{Bogolubskaja}, it is easy to see how "work"  the theory (\ref{3.17}) (say, in four-dimensional constraint-shell QED \cite{Pervush2, Azimov,Nguen2}) $A^D$, $\psi^D$, $\bar \psi^D$).
 
But  such a modification  does not affect the relativistic invariant S-matrix squared $\vert  S\vert ^2$ (that is  {\it on-shell } \rm 
of fields), invariant  with respect to the Heisenberg-Pauli  transformations
(\ref{ltf})- (\ref{lqed}) \cite{Heisenberg,Pervush2,Nguen2, Polubarinov} of the Dirac variables $A^D$, $\psi^D$, $\bar \psi^D$. 

But \it off-shell \rm   various spurious diagrams appear induced by "gauge constituents" $\Lambda$ in Heisenberg-Pauli  transformations
(\ref{ltf})- (\ref{lqed}). 

 The appearance of such  spurious diagrams in constraint-shell QED \cite{Pervush2, Azimov,Nguen2} is associated with the 
$\Lambda$-transformations (\ref {dsigLa}) \cite{Nguen2} for the  electronic Green function $G(p)$.
\section{Expanding the Dirac quantization scheme from QED to the Abelian $U(1)$ theory.}
The (Minkowskian) Abelian gauge model contains the  Abelian group $U(1)$, and this determines its nontrivial topological content:
\be \label {fundament}
\pi_1 (U(1))= \pi_1 S^1= {\bf Z},
\ee
in turn specified by the radius $\vert {\bf x}\vert <\infty$ of the circle $S^1$.

How this nontrivial topology (\ref{fundament}) is embodied in the Dirac fundamental quantization \cite{Dir} of the Abelian $U(1)$ model (with the exact $U(1)$ symmetry) we just attempt to elucidate in the present section.

\medskip The plan of this section is following. First, we shall demonstrate (although this is evident even on the face of it) that constraint-shell QED \cite{Pervush2, Azimov,Nguen2} is the topologically trivial sector ($n=0$) of the constraint-shell Abelian $U(1)$ model we construct now.

Secondly, we implement the Gauss-shell reduction of the Abelian $U(1)$ model with the exact $U(1)$ symmetry (further AM) for the nontrivial topologies ($n\neq 0$) involving the Dirac monopole modes \cite{Ryder,Dirac} and construct the Dirac variables for this case.

\bigskip Beginning with the point one of our programm, note that the three-dimensional configuration space $A^D=(A_0^T, A^D_i)$  ($i=1,2$) of Dirac variables in constraint-shell QED \cite{Pervush2, Azimov,Nguen2} is topologically equivalent  to the flat space ${\bf R}^3$ with the deleted origin of coordinates:  
\be \label{prokol}
A^D\simeq {\bf R}^3\setminus \{0 \}.
\ee 
This is connected closely with the manifest $O(1/r)$ behaviour of Dirac variables in constraint-shell QED, at which these nonlocal functionals of gauge fields are badly specified in the origin of coordinates (one can trace this behaviour of Dirac variables, for example, in Eq. (\ref{retard}) for retarding potentials \cite{Landau2} in the case of "plane waves",  depending explicitly on the  time $t$, as well as for  "electrostatic" solutions $A_k^D ({\bf x)}$ to the Poisson  equation (\ref{Poi}) and $A_0^T ({\bf x)}$ to the Poisson  equation in (\ref {QEDt})).

Ii is well known (see e.g. \S T1 in   \cite{Al.S.}) that
$$ {\bf R}^3\setminus \{0 \} \simeq S^2.$$ 
Whence 
\be \label {jeg} \pi_2 ({\bf R}^3\setminus \{0 \})= \pi_2 S^2={\bf Z}\neq 0.  \ee
Latter Eq. testifies in favour of the point hedgehog topological defect inside the manifold ${\bf R}^3\setminus \{0 \}$ in an infinitesimal neighbourhood of the  origin of coordinates (see \S$\Phi$1 in \cite{Al.S.}).

On the other hand, the fundamental homotopical group $\pi_1 ({\bf R}^3\setminus \{0 \})$ is 
\be \label{otop} \pi_1 ({\bf R}^3\setminus \{0 \})= \pi_1 S^2=0.\ee 
Eqs.  (\ref{jeg}), (\ref{otop}) imply, respectively, the point hedgehog topological defect inside the  configuration space $A^D$ of QED Dirac variables, in an infinitesimal neighbourhood of the  origin of coordinates, and the trivial fundamental homotopical group of this configuration space. 

In effect, the topological equality (\ref{otop}), compared with 
(\ref {fundament}), implies that the  QED Dirac variables (\ref {vak}) belong to the trivial topological sector of the  $U(1)$ group space, in spite of the bad definition of gauge fields $A^D$ at the  origin of coordinates.

\bigskip To perform the Gauss-shell reduction of AM, we should recall that in the basic of this model lies the {\it duality} \cite{Cheng} between the set of Maxwell equations
\bea \label{Max}
\nabla\cdot {\bf E}=\rho; \quad \nabla \times {\bf B}-\partial_0 {\bf E}={\bf j};\\
\nonumber
\nabla\cdot {\bf B}=0; \quad \nabla \times {\bf E} +\partial_0 {\bf B}=0,
\eea
written down as
$\partial_\mu F^{\mu \nu}=-j^\mu$ ($j^\mu=(\rho,{\bf j})$)
in terms of the Maxwell electromagnetic tensor $F^{\mu \nu}$, and the set of  equations
\be \label{dual}
\partial_\nu \tilde F^{\mu \nu}=-k^\mu; \quad  \tilde F^{\mu \nu}=\frac 1 2 \epsilon ^{\mu \nu\rho \sigma} F_{\rho \sigma}, \quad k^\mu=(\sigma,{\bf k})
\ee  for its dual $\tilde F$.

Otherwise (in 'classical' electrodynamics), it should be $\partial_\nu \tilde F^{\mu \nu}=0$ while $\partial_\mu F^{\mu \nu}=-j^\mu$ ('classical' electrodynamics is not symmetrical with respect to the intercharge of the electrical, $\bf E$, and magnetical, $\bf B$, tensities \cite{Cheng}: ${\bf E}\to{\bf B}$, ${\bf B}\to -{\bf E}$).

It is apropriate, in this point, to introduce (as it was done already, see eg. \cite{Fry}) the tensor
\be \label{sum tens}
{\cal F}^{\mu \nu} =F^{\mu \nu}+\tilde F^{\mu \nu}
\ee
for which the {\it Cabibbo-Ferrari-Shanmugadhasan relation}
\be \label{Cab}
{\cal F}^{\mu \nu} =\partial^\mu A^\nu-\partial^\nu A^\mu +\epsilon ^{\mu \nu\rho\sigma}\partial_\rho \tilde A_\sigma
\ee
takes place \footnote{In this context the potential $\tilde A^\mu$ concerns the dual tensor $\tilde F^{\mu \nu}$ in the same wise as the potential $ A^\mu$ concerns the Maxwell tensor $ F^{\mu \nu}$. But below, when I will talk about magnetic monopoles, distracted somehow from electrmagnetic field $ F^{\mu \nu}$, I will utilize merely the symbol $A^\mu$ for such potential, associated with magnetic monopoles.}.

\medskip The {\it magnetical current} $k^\mu=(\sigma,{\bf k})$ introduced in (\ref{dual}) saves the situation, i.e. it restores the above ${\bf E}\to {\bf B}$; ${\bf B}\to -{\bf E} $ (or $F^{\mu \nu}\to \tilde F^{\mu \nu}$; $\tilde F^{\mu \nu}\to -F^{\mu \nu}$) symmetry.

Herewith Eqs. (\ref{Max}) and (\ref{dual}) imply also the duality transformations
\be \label{toki}
j^\mu\to k^\mu; \quad k^\mu\to-j^\mu 
\ee
for the currents.

\medskip Since the electromagnetic current $j^\mu $ is present in AM, we now consider, the Gauss law constraint (\ref{Gau}), with its solution of the shape (\ref{var1s}), remains valid in this model. Once again, $A_0$ proves to be the nondynamical degree of freedom, and now we attemt to remove it with the aid of some gauge transformations, the shape of which we shal elucidate below.

\bigskip Following \cite{Cheng}, let us consider the Schr${\rm \ddot o}$dinger equation for a fermion in the electromagnetic background $(A_0,{\bf A})$. This is 
\be \label{Schrod}
[\frac 1{2m}({\bf p}-e{\bf A})^2 +eA_0]\psi=i \frac {\partial \psi}{\partial t},
\ee
 invariant with respect to the gauge transformations
 \bea \label{gt1}
 {\bf A}({\bf x})\to {\bf A}({\bf x})+ \frac 1 e { \nabla} \alpha({\bf x});\\
 \nonumber
 \psi({\bf x})\to e^{i\alpha({\bf x}}\psi({\bf x})
 \eea
 
\medskip The situation with the  Schr${\rm \ddot o}$dinger equation (\ref{Schrod}) is "good", i.e. this {\it does not has singular solutions}, when the magnetic current $k^\mu$ is "switched off". But if the magnetic current $k^\mu$ is "switched on", the vector potential $\bf A$ {\it cannot exist everywhere}, proving to be singular at definite values of $\bf x$. Now we attempt to find out why it is so, what these values of $\bf x$ are and how to avoid this difficulty.

Note firstly that the solution to the duality equation (\ref{dual}) is the potential magnetic field \cite{Cheng,Ryder}
\be \label{radial}
{\bf B}=\frac g{4\pi r^2}{\bf n}; \quad {\bf n}=\frac {\bf r}{r}; \quad {\rm div} {\bf B}=4\pi g\delta^3({\bf r}).
\ee
Here $g$ is the {\it magnetic charge} \cite{Cheng,Ryder,Dirac}, connected with the $\sigma\equiv k^0$ component of $k^\mu$ by the relation \cite{Cheng}
\be \label{deltaf}
k^0=\sum_i g_i\int dx_i^0 \delta^4(x-x_i).
\ee
Whence 
\be \label{circ}
4\pi g=\oint_S {\bf B}\cdot d{\bf S}.
\ee
It is obvious that $\bf B$ cannot be written down as $\nabla \times {\bf A}$, then $${\rm div}(\nabla \times {\bf A})={\rm div}{\bf B}=0$$
and the integral (\ref{circ}) becomes zero: in this case no magnetic charges $g$ exist (we are not interested in this trivial case now). However, one can define $\bf A$ in such a wise that $\bf B$ is given as $\nabla \times {\bf A}$ {\it everywhere except on a line joining the origin of coordinates to infinity}. To see that it is possible, let us consider \cite{Cheng} the magnetic field created by the infinitely long and thin solenoid placed along the negative $z$-axis with its positive pole (with the strength $g$) at the origin:
\be \label{solen}
{\bf B}_{\rm sol}=\frac g{4\pi r^2}{\bf n} +g\theta (-z) \delta(x)\delta(y)\hat {\bf z},
\ee
where $\hat {\bf z}$ is the unit vector in the $z$-direction. This magnetic field differs from the magnetic monopole field (\ref{radial}) by the singular magnetic flux along the solenoid, set by the second item in (\ref{solen}). Sinse the magnetic field given in (\ref{solen}) is source-free ($\nabla\cdot {\bf B}_{\rm sol}=0$), one can write
\be \label{solen1}
{\bf B}_{\rm sol}=\nabla\times {\bf A}.
\ee
Then from (\ref{radial}), (\ref{solen}), (\ref{solen1}) one derives the monopole field given as
\be \label{radial1}
{\bf B}=\frac g{4\pi r^2}{\bf n}=\nabla\times {\bf A}-g\theta (-z) \delta(x)\delta(y)\hat {\bf z}.
\ee
The line occupied by the solenoid is called the {\it Dirac string}. The potential $\bf A$, enering Eq. (\ref{radial1}), singular along the negative $z$-axis, can be set as \cite{Ryder}
\be \label{adec}
A_x=g \frac{-y}{r(r+z)}; \quad A_y=g \frac{x}{r(r+z)}; \quad A_z=0;
\ee
or
\be \label{acil}
A_r=A_\theta=0;  \quad A_\phi=\frac g r \frac{1-\cos \theta}{\sin \theta}
\ee
in the spherical coordinates.

\medskip The Dirac string is a purely gauge artefact. So, for instance, if the Dirac string is located along the line $r=z$, instead of Eq. (\ref{acil}) one should write \cite{Ryder}
\be \label{acil1}
A_r=A_\theta=0;  \quad A_\phi=-\frac g r \frac{1+\cos \theta}{\sin \theta}.
\ee
The only {\it physical} singularity of the potential $\bf A$ is its  singularity at the origin of coordinates $r=0$. This shows clearly that the Dirac monopole (\ref{radial1}) is an example of point (hedgehog) topological defects (see e.g. \S$\Phi$1 in \cite{Al.S.} and the discussion in \cite{disc}). 

\bigskip Now we get down directly to constructing Dirac variables in AM. 

To do this, let us study \cite{Ryder} the quantum behaviour of a charged particle (with the elementary charge $e$) in the magnetic monopole field. Its wave function is
\be \label{wave}
\psi=\vert\psi\vert \exp [\frac i\hbar ({\bf p\cdot r}-Et)].
\ee
When the magnetic monopole field is "switched on", we have ${\bf p}\to {\bf p}-(e/c){\bf A}$ (where $\bf A$ is given in (\ref{adec}) or in (\ref{acil})), and
$$\psi\to \psi \exp(-\frac{ie}{\hbar c} {\bf A\cdot r}),$$
i.e. the change in the phase $\alpha$ of the wave function (\ref{wave}),
\be \label{change}
\alpha\to\alpha-\frac e {\hbar c}{\bf A\cdot r}
\ee
occurs.

Let us consider now a closed contour at the fixed $r$, $\theta$, $\phi \in[0,2\pi]$. Then the complete change in the phase $\alpha$ will be
$$ \label{complete change}
\Delta\alpha =\frac e {\hbar c} \oint {\bf A} \cdot d{\bf l}=\frac e {\hbar c} \int {\rm rot}{\bf A}\cdot d{\bf S}=\frac e {\hbar c} \int {\bf B}\cdot d{\bf S}=$$ 
\be \label{completechange}
\frac e {\hbar c}({\rm the~ flux~ across~ the~ part~ of~ the~sphere })=\frac e {\hbar c} \Phi(r,\theta),
\ee
where $\Phi(r,\theta)$ is the flux across the part of the sphere specefied by some values of $r$ and $\theta$. As $\theta$ is changed,  the flux across this part of the sphere is also changed. So, as $\theta\to 0$, the contour is shrinked to a point, and the flux passing across this part of the sphere goes to zero, $\Phi(r,0)=0$.

As the contour is increased, the flux is also  increased, and at last as $\theta\to\pi$,
\be \label{potok}
\Phi(r,\pi)=\int {\bf B}\cdot d{\bf S}=4\pi r^2B=4\pi g
\ee
due to (\ref{radial}).

But since at $\theta\to\pi$ {\it the contour again is shrinked to a point}, the potential $\bf A$ {\it should be singular} at $\theta=\pi$ in order for $\Phi(r,\theta)$ to be finite. And 
moreover, this conclusion is correct at any value of $r$, i.e. for a sphere of any radius; thus the potential $\bf A$ is singular along the negative half-axis $z$.
This is an alternative deriving the above results  (\ref{adec}),  (\ref{acil}). It is obvious herewith that the Dirac string can be located along any direction (and, generally speaking, it is not  definitely that it is a straight line, but it should be contineous; in other words, it should be a Jordan curve).

Note that the discussed singularity of the potential $\bf A$ implies the so-called "Dirac veto" \cite{Ryder}: the wave function $\psi$ goes to zero on the negative half-axis $z$. Therefore, its phase along this line is not specified and it follows from Eq. (\ref{completechange}) that there are no necessity in the condition $\Delta \alpha\to 0$ at $\theta \to \pi$. But $\psi$ should be an uniquely defined function; thus the equality 
$$\Delta \alpha=2\pi n, \quad n \in {\bf Z},$$
should be satisfied.

Then from (\ref{completechange}), (\ref{potok}) one derives
$$2\pi n=\frac e {\hbar c} 4\pi g$$
or
\be \label{dq}
eg=\frac 1 2 n \hbar c
\ee
(note that it is precisely Eq. (\ref{qq}) in the $\hbar = c=1$ system of units; here is $g$ instead of $\bf m$ stands for the magnetic charge).

 \bigskip The evident dualism between the radial magnetic field (\ref{radial}) and the  tension ${\bf E}$ in usual electrostatics suggests the "Coulomb-like" behaviour of the force between two magnetic charges $g_1$ and $g_2$, just 
 \be \label{mCoulomb}
 F=\frac {g_1g_2}{r^2}.
 \ee
 It can be also concluded that two  magnetic charges of different signs are attracted while those with same sign  repel. It is easy to see herewith that the "algebra" of magnetic charges coincides with that of integer numbers $\bf Z$.

\bigskip Now the look for the Dirac variables in AM can be derived by analogy with (\ref{vak}) \cite{Pervush2}, involving the gauge factor $v[A]$, (\ref{va1}). Herewith the finiteness of the phase $\Phi(r,\theta)$, (\ref{completechange}), should be taken into account. Also it is important that now the gauge potential $\bf A$ is stationary \footnote{For instance, there are no electromagnetic fields bearing non zero topological numbers,{ \it Nature does not know merely such fields}! Below wee shall also 
clarify what occurs in the "zero" topological sector of the AM model.
 },  generating the Dirac monopole; thus instead of  (\ref{va1}) it should be a stationary multiplier, as it was in the non-Abelian YMH model (quantized by Dirac), involving vacuum BPS monopole modes (us discussed briefly in Introduction). 

Meanwhile, if the gauge potential $\bf A$ is given by Eqs. (\ref{acil}) or (\ref{acil1}) \cite{Ryder}, the Gauss law constraint  (\ref{var1}) remains formally valid, but now its r.h.s. becomes 
\be \label{Gauss m}
\Delta A_0=j_0
\ee
if the fermionic current $j_0$ exists. But if the latter one is "turned off", we encounter the simple homogeneous Laplace equation 
\be \label{Gauss m1}
\Delta A_0=0,
\ee
permitting the above citing solution (\ref{homogen}) \cite{Nguen2}.



On the other hand, now (when the duality transformations (\ref{toki}) are performed, implyig the Dirac conjucture (\ref{dq}) \cite{Dirac} is correct) there is no any sense to set $A_0=0$ for nontrivial topologies $n\neq 0$ (identically or by performing some transformations). Now we shell attempt to ground this.

If $A_0=0$, the electric tensity 
$$\bf E= \partial{\bf A}/\partial t +\nabla A_0$$
turns into a zero field over the stationary monopole configuration (\ref{adec}) (when another vector potentials: say, electromagnetic, are absent). This means the absence of (stationary) electric charges $Qe \equiv Q$ (at setting $e=1$) \footnote{Instead of the Dirac quantization conjecture \cite{Dirac} given in shape (\ref{dq}), one utilize very often its shape 
$$\frac{Qg}{4\pi}= \frac 1 2 n.$$
 Grounding the latter equation is given, for instance, in the monograph  \cite{Cheng} in \S 15.1).
} (which serve, if exist, as  sourses of [electrostatic] fields) in the model. According Eq. (\ref{dq}) this implies that only the "especial" case of the trivial topology $n=0$ has right to exist without of essencial problems; otherwise one encounter infinite magnetic charges $g$,  and then the magnetic field $\bf B$ can be  finite/infinite (instead of to be {\it zero}) at the spatial infinity (this is according to (\ref{radial})).  This, in turn, can lead to the infinite energy integral 
$$ \sim  \int d^4x (\tilde F^{\mu \nu})^2$$
 as a consequence.
 
 And also all this implies the bad renormgroup properties of the theory in question and divergences in apropriate Feynman diagrams involving different degrees of $g$. And vice versa, the 
chance to get a good perturbation theory appears  apparently at including (at nontrivial topologies) Feynman diagrams with $g$ as well as with $e$.  Likely, in this case, a fine interplay between these diagrams implies a cancelation of divergences.

 \medskip Thus if we desire to preserve nontrivial topologies $n\neq 0$ in our Abelian $U(1)$ model, we should rule out the gauges in which $A_0=0$ (the "especial" case $n=0$ in which only magnetic charges can exist is, indeed, an intersting case, we shall discuss below).
 
 This means that "turning off" the charge $j_0$ in the Gauss law constraint (\ref{Gauss m}) has no physical sense in the us discussed theory, generally speaking (except  the trivial topology case $n=0$).
 
 In the said is the specific of stationary monopole solutions (\ref{adec}) appearing in the "complete" (topologically nontrivial) $U(1)$ gauge model in which the  Dirac conjucture (\ref{dq}) is assumed. In this is the essential difference of this model from  "pure" constaint-shell QED (\ref{LEDt}) \cite{Nguen2} permitting solutions (Dirac variables) depending manifestly on time.
 
 \medskip Now we can begin directly with writting down Dirac variables in the above topologically nontrivial $U(1)$ gauge model involving the monopole configuration (\ref{adec}). The experience of constaint-shell QED \cite{Azimov,Nguen2} and Minkowskian YMH model with BPS monopoles quantized by Dirac \cite{fund} allows us to do this.
 
 First of all, it is easy to see that the analogue of the Gribov phase $\hat\Phi_0({\bf x})$ in the Minkowskian YMH model with BPS monopoles quantized by Dirac, in the $U(1)$ gauge model with Dirac monopoles will be the value $\Phi (r,\theta)$ \cite{Ryder}. This can be interpreted as a stationary phase for (topological) Dirac variables in the investigated $U(1)$ gauge model involving Dirac monopole modes. 
 
 As a result, by analogy with $\hat A^{(n)}_k$ in the non-Abelian gauge model, in our case the apropriate (topological) Dirac variables can be written down as
 
 \be \label{dva}
 A^{Dm (n)}_\mu=v^{Dm(n)}({\bf x}) ({ A}_\mu^{ (0)}+
\partial _\mu)v^{Dm(n)}({\bf x})^{-1},\quad v^{Dm(n)}({\bf x})=
\exp [n\Phi (r,\theta)]; \quad \mu=0,1,2,3.
 \ee
 Here ${ A}_\mu^{ (0)}$ is the topologically trivial field configuration (in which the electric charge takes arbitrary values while the magnetic charge of such a configuration is zero). From Eqs. (\ref{adec}) -(\ref{acil1}) it can be concluded that the spatial components $A^{Dm (0)}_i$ ($i=1,2,3$) are exactly zero since now $g=0$ {\it is fixed}, even in the singularity point $r\to 0$. But their topological copies  $A^{Dm (n)}_i$
 in another sectors can be different from zero due to the presence of the value $\Phi (r,\theta)$ in (\ref{dva}) \footnote{Such look for the topological multipliers $v^{Dm(n)}({\bf x})$ as in (\ref{dva}) is not a qiite mandatory because  $\Phi (r,\theta)$ is a function of the topological number $n$:  this is according to Eqs. (\ref{potok}) , (\ref{dq}). So, indeed, $\Phi (r,\theta)\equiv \Phi_1 (r,\theta, n)$ and the look $v^{Dm(n)}({\bf x})=\exp [\Phi_1 (r,\theta, n)]$ for the above topological multipliers $v^{Dm(n)}({\bf x})$ is quite acceptable.}.
 
 On the other hand, since Dirac variables in any model are gauge invariant inherently, the transformation (\ref{dva}) can be treated as a map from the zero to the $n$$^{\rm th}$, nontrivial, topological sector of the model discussed. {\it But it is not a gauge transformation}!  The same applies to Dirac variables in any model with nontrivial topologies, for instance, the non-Abelian YMH model  \cite{LP1,David3,LP2, Pervush2,  David1}.  In that case the "large" Dirac variables $\hat A_k^{D(n)}$ is the image of "small" $\hat A^{(0)}_k$ at the topological map of such kind.
 
 As it was noted in \cite{rem2}, there exists a simple mathematic model (see Lecture 2 in \cite{Postn4}) which  describes correctly such a topological map. In the studied case, the base of the covering consists of all the topologically trivial gauge fields $A_\mu^{(0)}$, while its discrete infinitely-valent fibre is the set of all fields $A^{Dm (n)}_\mu$.
 
  \medskip As regards the temporal components $A^{Dm (n)}_0$ of the Dirac variables (\ref{dva}), these ({\it different from zero} due to the above reasoning), as it is easy to understand, satisfy the Gauss law constraint (\ref{Gauss m}), that is the Poisson equation, permitting  purely stationary (${\rm O}(1/r)$) solutions. Thus they are associated with electric charges inherent in model we study now. And moreover, as it can be seen from  Eq. (\ref{dva}) for a potential $A_0$, the topological copies with $n\neq 0$ for the topologically trivial temporal potential ${ A}_0^{ (0)}$ {\it coincide}  with this ${ A}_0^{ (0)}$ since we deal in the considered model with the topological multipliers $v^{Dm(n)}({\bf x})$ which are commute and are independent on $x_0$.
 
 The proof that these variables (\ref{dva}) are gauge invariant is similar to that \cite{Azimov} in constraint-shell QED (see the calculations (\ref{proverca})), with only constructive addition that now there is no 
necessity in the nonstationary gauge matrices (such as $g$, (\ref{gm}) in constraint-shell QED). Now the gauge matrices can be chosen to be purely stationary, say 
\be \label{newg}
\tilde g({\bf x})\equiv \exp (ie\Lambda ({\bf x}))
 \ee
 Also, formally, the condition $\partial_\mu \partial^\mu \Lambda =0$ can be imposed onto $\Lambda ({\bf x})$.
 
 \medskip In contrast to the Dirac variables in  constraint-shell QED or in the Minkowskian YMH model with BPS monopoles quantized by Dirac, the Dirac variables (\ref{dva}) in the  $U(1)$ gauge model with Dirac monopoles are not transverse automatically but such condition can be imposed onto these variables. This involves some 
cumbersome 
mathematics affecting the monopole configuration (\ref{adec})-  (\ref{acil1}) \cite{Ryder} and the "Gribov phase" $\Phi (r,\theta)$. We  omitt them in the present study.
 
 \medskip Note that (\ref{dva}) is the purely stationary field configuration, in contrast to $\hat A_k^D$ in the Minkowskian YMH model with BPS monopoles quantized by Dirac.
 
 \medskip Now (upon grounding the gauge invariance of the Dirac variables $A^{Dm (n)}_\mu$), it becomes obvious that the gauge $A_0^{Dm(0)}=0$ (can exist in the unique case of arbitrary magnertic charges $g$, zero electric charges and the zero topology $n=0$) gives the gauge $A_0^{Dm(n)}=0$ in all the nontrivial  topological sectors: it is according to Eq. (\ref{dva}). But such topologically nontrivial field configurations should be ruled out as those can lead to the infinite Hamiltonian (Lagrangian) density because of the now becoming infinite magnertic charges $g\sim n/\infty$. 
 
 Due to the gauge invariance principle for the Dirac variables $A^{Dm (n)}_\mu$, (\ref{dva}), we can write down for the model Hamiltonian $H$:
 \be \label{gi}
 H(A_0^{Dm(0)})=H(A_0^{Dm(n)})
 \ee
 The latter equation implies ruling out also the  topologically trivial field configuration $A_0^{Dm(0)}$. Thus one can suppose that {\it purely magnetic} and electrically neutral particles, creating the "radial" magnetic field $\bf B$ according to Eq. (\ref{radial}) (such particles can be referred as "magnons"), {\it can not exist, probably, in the Nature}.
 
 \bigskip Indeed, as we'll argue now, the manifest gauge invariance (\ref{gi}) of the model Hamiltonian $H$ prohibits, in general, the existense of magnetic charges in  the Nature if the look (\ref{dva}) for the topological Dirac variables in the $U(1)$ gauge theory is assumed.
 
 Really, from Eq. (\ref{adec}) it follows that in the zero topological sector at an arbitrary electric charge $Q\neq 0$ (then $g=0$ due to (\ref{dq})), the vector potential $\bf A$ \cite{Cheng,Ryder}, (\ref{adec}), generating the magnetic monopole solution (\ref{radial1}), becomes zero (even in the physical singularity point ${\bf r}\to 0$ since $g=0$ is fixed).
 
 Then the  gauge invariance  of the model Hamiltonian, $H({\bf A}^{Dm(0)})=H({\bf A}^{Dm(n)})$, implies, as it is easy to see, {\it the impossibility to observe magnetic charges} (but only  electric charges!) if the shape (\ref{dva}) for the topological Dirac variables  is assumed. The situation remind us that in the non-Abelian YMH gauge model (with vacuum BPS monopoles) quantized by Dirac.  As it was argued in Ref.  \cite{Pervush2},  due to the  gauge invariance,  $H[A^{(n)},q^{(n)}]=H[A^{(0)},q^{(0)}]$,  the QCD Hamiltonian $H$ does not depend on the Gribov phase factors $v^{(n)}({\bf x})$ and "it contains the perturbation series
in terms of only the zero  map fields (i.e., in terms of constituent color particles) that can be
identified with the Feynman partons" \cite{Pervush2} (in other words, only topologically trivial quarks $q^{(0)}$ and gluons $A^{(0)}$ can be observed). 

\bigskip  In order to maintain in these circumstances the Dirac conjecture (\ref{dq}) \cite{Dirac} about the quantization of the magnetic charge, the following way out seems to be quite reasonable. We propose to construct the topological (Gribov) copies of the potentials (\ref{adec}) containing manifestly the magnetic charge $g$ and generating the Dirac monopole given in Eq. (\ref{radial1}). Note herewith that alone the {\it gauge covariant} potentials (\ref{adec}) are topologically nontrivial sinse they contain $g$ manifestly and then depend on the apropriate topological number $n$ via Eq. (\ref{dq}) \cite{Dirac}.

By analogy with (\ref{dva}),  we write now
\be \label{dvn}
A_{i}^{'Dm(n)}=v^{Dm(n)}({\bf x}) (A_i^{(n)}+ \partial_i)v^{Dm(n)}({\bf x})^{-1}; \quad i=1,2,3,
\ee
where now $A_i^{(n)}$ are the spatial potentials given in Eq. (\ref{adec}). 

The same "transformation law" can be written down also for the temporal components of the potentials $A$: {\it separately in each topological sector of the considered model}:
\be \label{dvn0}
A_{0}^{'Dm(n)}=v^{Dm(n)}({\bf x}) (A_0^{(n)}+ \partial_0)v^{Dm(n)}({\bf x})^{-1}=A_0^{(n)}
\ee
since the topological multipliers $v^{Dm(n)}({\bf x})$, (\ref{dva}), are the Abelian multipliers which commute with $A_0^{(n)}$ and due to their explicit look (they are manifestly stationary).

Note that setting $A_0^{(n)}=0$ for any $n\neq 0$ in (\ref{dvn0}) implies $Q(n)=0$ for this  topology $n$ and then the infinite magnetic charges $g(n)\to n/\infty$ in this  topological sector due to Eq. (\ref{dq}) \cite{Dirac}. As it was explained above, this can lead to the apropriate infinite energy integral and another undesirable consequences. Thus  setting $A_0^{(n)}=0$ is 
not profitable for us. 

This has a very important consequence. At assuming  $A_0^{(n)}=0$ ($n\neq 0$), one concludes easly that plane transverse vawes (photons on the quantum level) are {\it impossible} for nontrivial topologies ($n\neq 0$) in the Abelian gauge $U(1)$ model involving magnetic charges. Such solutions, as it is well known (for example, \S 36 in \cite{Landau2}), must presume just the Coulomb gauge $A_0^{(n)}=0$ ($n\neq 0$). Thus, {\it topologically nontrivial photons do not exist}  in the Abelian gauge $U(1)$ model involving magnetic charges.

\bigskip Now it is possible to write down explicitly the Hamiltonian for the $U(1)$ gauge model involving magnetic monopoles now taking into account the  transformation laws (\ref{dvn}), (\ref{dvn0}). Such {\it gauge invariant} Hamiltonian is, indeed, a sum, running through all the topologies $n\in {\bf Z}$, of the Hamiltonians  $H^{(n)}(A_i^{'Dm(n)}, A_0^{(n))}$:
\be \label{Ham}
{\cal H}=\sum _n H^{(n)}(A_i^{'Dm(n)}, A_0^{(n))}).
 \ee
 On the other hand, the  gauge invariance of the Hamiltonian $\cal H$ implies
\be \label{Ham1}
{\cal H}=\sum _n H^{(n)}(A_i^{(n)}, A_0^{(n))}),
\ee
where $A_i^{(n)}$ are the spatial potentials given in Eq. (\ref{adec}) and generating the Dirac monopoles \footnote{In this summing by topologies in the Hamiltanian $\cal H$ and also in fixing topological variables $A_{i}^{'Dm(n)}$, (\ref{dvn}), in each topological sector of the $U(1)$ Abelian model is the 
principled 
difference of this model from the quantized by Dirac YMH non-Abelian model \cite{LP1,David3,LP2, Pervush2,  David1} with vacuum BPS monopole solutions. Here we are interested in fact only in the topologically trivial and  gauge invariant Hamiltanian $H[A^{(0)},q^{(0)}]$, describing correctly the quark confinement \cite{Pervush2}.}.

\medskip The notable feature of the quantization scheme (\ref{dvn}), (\ref{dvn0}) is that it {\it permits} (unlike the quantization scheme (\ref{dva})) the existence of magnons (purely magnetically charges particles) in the zero topological sector of the considered model when $ A_0^{(0))}=0$ and the magnetic charge $g$ takes arbitrary, different from zero, values.

And vice verse, $g=0$ corresponds to the trivial topology $n=0$ with only arbitrary electric charges $Q$ existing. This is just the electrostatics case when these electric charges $Q$ induce the nonzero (electrostatic) potentials $ A_0^{(0))}\neq 0$. 

\medskip It is easy to see now that  "pure" constraint-shell QED (\ref{LEDt}) \cite{Nguen1}, in which the electric charges are absent and which is the direct result of the removal (\ref{udalenie1}) of the temporal potentials of electromagnetic potentials, 
supplements the topologically trivial "magnon sector" of the generalized $U(1)$ gauge model (involving Dirac monopoles). The same is correctly also for transverse electromagnetic waves (photons).

And 
at last, note that   constraint-shell QED (\ref{QEDt1}), involving fermionic currents, can be considered as the topologically trivial part of the mentioned generalized $U(1)$ gauge model in which $g=0$ is set. Herewith the  "temporal"' potentials $ A_0^{(0))}\neq 0$ are set by means of Eq. (\ref{dub}) \cite{Nguen1} while the "`spatial" (retarding) potentials  ${\bf A}^{D}(x)$, (\ref{retard}), again supplement the former.

The said is in a good agreement with our "topological" survey (\ref{prokol})- (\ref{otop}) grounding that the Dirac variables (\ref{vak}) \cite{Pervush2} belong to the zero topological sector of the $U(1)$ group space \footnote{Indeed, the topology (\ref{prokol})- (\ref{otop}) relates equally to the Dirac potentials $A^{D}=(A^{T}_0, A_i^{D})$ as wel as to the
conventional 
 Coulomb-like electrostatic potentials $\phi\equiv A_0$, obeying the Coulomb law in the shape  \cite{Al.S.}
 $$ {\rm div} ~{\rm grad}~\phi ={\rm div} ~{\bf E}\neq 0$$
 and also to the magnetic monopole field ${\bf B}=\tilde F^{\mu \nu}$ (generating the differential 2-form $\Omega=\tilde F^{\mu \nu} dx_\mu \wedge dx_\nu$).
 
 The important point in the both cases is the singularity of the physical fields at the origin of coordinates generating the topology (\ref{jeg}) (for details see $\S$ T7 (p. 278) in the monograph \cite{Al.S.} or p. 653 in the monograph \cite{Dubrovin}).
 }.

\medskip We finish our investigation of the Hamiltonian (\ref{Ham1}) by writing it explicitly. Note firstly that the  Lagrangian density for the bosonic sector of the us investigated model has the look \cite{Fry} 
\be \label{Lagu}
{\cal L}_{\rm bos}=-\frac 1{16\pi} {\cal F}^{\mu \nu} {\cal F}_{\mu \nu}
\ee
according to (\ref{Cab}). Herewith in the product of two tensors ${\cal F}^{\mu \nu}$ we hold only the terms ${\tilde F}^{\mu \nu} {\tilde F}_{\mu \nu}$ and ${ F}^{\mu \nu} { F}_{\mu \nu}$. The reasoning here is in  
neglecting (on the classical level) the interaction between the Maxwell electromagnetic field ${ F}_{\mu \nu}$ and its dual,   ${\bf B}\equiv{\tilde F}^{\mu \nu}$ (that is the magnetic monopoles configuration).  Such interaction refers to the quantum corrections of the fourth order (more exactly ${\rm O}(ge)^4$), by analogy with the photon-photon  scattering process in QED, involving four (virtual) fermions (see \S 41.1 in  \cite{A.I.}). 

Since $  A^{ \tilde \dot i(n)}=0$ ($i=1,2,3$) \footnote{Now we write down the symbol "tilde" over $A$ in order to distinguish the "Maxwell" and its dual tensors.} for the stationary configuration (\ref{adec}) (or (\ref{acil})) and since the canonical momenta
$$ \tilde p_0^{n}=\frac{\partial {\cal L_{\rm bos}}}{\partial A_0^{{\tilde \dot}(n)}}=0; \quad n\neq 0$$
(and the same is correct for the canonical momentum $p_0$ conjugate to $A_0$; as a result,  the Gessian matrix $M$ of the investigated $U(1)$ Abelian gauge model is again degenerate, as in ordinary QED, for instance) \cite{Gitman},
one can write down for the Hamiltonian ${\cal H}$: 
\be \label{Ham11}
{\cal H}= p_0\dot A_0+ \sum _{n\neq 0} \tilde p_0^{n}A_0^{{\tilde \dot}(n)}+ (F^{i0})^2+\tilde p_i^{n} A^{ \tilde \dot i(n)}-{\cal L}_{\rm bos}.
\ee
Here, obviously, \cite{Gitman}
$$ F^{i0}=E^{i}=  p_i^{0}.$$ 
All this results
\be \label{hne0}
H^{(n)}(\tilde A_i^{(n)},\tilde A_0^{(n)})=\frac 1 {16 \pi} \tilde F_{ij (n)}\tilde F^{ij (n)}; \quad n\neq 0,
\ee
where $\tilde F_{ij (n)}$ is the part of the tension tensor $\tilde F$ corresponding to the topology $n$.

If $n=0$,
\be \label{hr0}
H^{(0)}(A_i^{(0)},A_0^{(0)},\tilde A_i^{(0)})=\frac 1 {16 \pi}  F_{ij } F^{ij } +\frac 1 {8 \pi}F_{i0}F^{i0} +\frac 1 {16 \pi} \tilde F_{ij (0)}\tilde F^{ij (0)}.
\ee
The latter term in (\ref{hr0}) takes account of the "magnon" contribution (arbitrary $g$ at zero $Q$) into the zero topological sector of the discussed model, while first two describe correctly the "purely Maxwell theory" (in particular, QED). As for the mixed items $\sim F_{ij }\tilde F^{ij (0)}$ or $F_{i0}\tilde F^{ij (0)}$ these give a disapearing contribution into the Hamiltonian $H^{(0)}(A_i^{(0)},A_0^{(0)},\tilde A_i^{(0)})$
due to the reasoning given above.

 
 
 \bigskip  Now let us analyse the  fermionic sector of the Abelian $U(1)$ gauge model involving Dirac monopoles and how to incorporate the Dirac variables into this sector.
 
  \medskip The question about the  Dirac variables in the fermionic sector of the "complete" $U(1)$ gauge model is very interesting and important. In the light of the "dyon/magnon picture",  we assume the shape 
  \be \label{mag.tok}
 k_\mu=g \psi \gamma _\mu \bar \psi,
  \ee
  with $\gamma _\mu$ being the usual Dirac marices and $\psi$, $\bar \psi=\gamma _0 \psi$ being the fermionic field possesing the $1/2$ spin 
 for the magnetic current $k_\mu$ (which describe correctly magnons as well as dyons depending on the topological number $n$).
 
 This assumption is quite natural and obvious sinse the Dirac conjecture \cite{Dirac} does not affect the Lorentz and spinoral properties of values involved in the $U(1)$ gauge model with Dirac monopoles.
 
 \bigskip The "technology" writing down Dirac variables in  the  fermionic sector of the Abelian $U(1)$ gauge model involving Dirac monopoles is the same as in the gauge fields case discussed above (and also we have the pattern how to write down the Dirac variables in constraint-shell QED, see Section 2)
 
So we write down

\be \label{ferd}
\psi^{'(n)}=v^{Dm(n)}\psi_0^{(n)},
\ee
with $\psi_0^{(n)}$ being the "initial", {\it topologically nontrivial}, data for the fermionic field $\psi$, in the $n^{\rm th}$ sector of the $U(1)$
gauge model. Herewith it is important to note that the field $\psi_0^{(n)}$ is {\it gauge covariant} while the Dirac variable $\psi^{'(n)}$ is {\it gauge invariant}.
 
 \bigskip Since the both electric and magnetic charges should always be present in the us discussed "complete" $U(1)$ gauge model, this, by analogy with QED, allows us now to write down the model Hamiltonian. This should have the shape 
 
  \bea \label{Hamil} 
  H=(\sum_n H^{(n)})-i\bar \psi_0^{(0)} \gamma^i(\partial_i-ie A^{(0)}_i -ig\tilde A_i^{(0)})\psi_0^{(0)} -i\sum_{n\neq 0}\bar \psi_0^{(n)} \gamma^i(\partial_i-ig\tilde A_i^{(n)})\psi_0^{(n)}-\nonumber \\i\bar \psi_0^{(0)} \gamma^0 (\partial_0-ie A^{(0)}_0)\psi_0^{(0)}- \nonumber \\ i\sum_{n\neq 0}\bar \psi_0^{(n)} \gamma^0(\partial_0-ig A^{(n)}_0)\psi_0^{(n)} + M(\psi,\bar \psi) + \quad {\rm mixed~~ items} \nonumber \\
  \eea 
  Here $\sum _n H^{(n)}$ stands for the complete "bosonic" Hamiltonian including the items (\ref{hne0}), (\ref{hr0}) and $M(\psi,\bar \psi)$ is the fermionic mass item. The remarkable point of Eq. (\ref{Hamil}) is also that $A^{(0)}_0$ are the electrostatic potentials, as it was argued above. Also we note that only "initial" values $\psi_0^{(n)}$ ($n\in \bf Z$) of fermionic fields enter the Hamiltonian (\ref{Hamil}) due to its manifest gauge invariance.
  
  The "mixed items" in the last equation are very important and have the following meanings. There are the items describing the interactions between the fermions (including the coulombic one)  via gauge fields belonging to different topological sectors of the considered model.

  \bigskip  Concluding this section, the author should like express his opinion about the following objection against existing magnetic charges (and magnetic monopoles, as a consequence) in Nature. 
  
  So, for instance, in the paper \cite{tom}, it was argued that the  transformations 
  
  \bea \label{tt}
  j _\mu\to j _\mu \cos \theta +k _\mu\sin \theta;\nonumber \\
  k_\mu\to -j_\mu \sin \theta +k _\mu\cos \theta
  \eea
  for the electric/magnetic currents leave invariant the equations of motion (\ref {dual}),  the Lorentz force
  \be\label{lf}
  f^\nu =j _\mu F^{\mu \nu}+ k _\mu \tilde F^{\mu \nu}
  \ee
  and also the appropriate energy-momentum tensor for the electromagnetic field
  if a dyon is involved.
  
  Thus the question about the parameter $\theta$, entering (\ref{tt}), with fixing its concrete value  is rather a matter of convention but not of an experimental choice.
  
  If now one considers a totality of dyons for wich the $g/Qe$ ratio has the same arbitrary chosen value, then the above parameter  $\theta$ can be connected with this ratio, for insance, as 
  \be \label{arc}
  \theta =\arctan (g/Qe).
  \ee
  Then with the aid of the dual rotations
  
  \bea
  \label{dual povorot}
  {\cal F}_ {\mu \nu}=(Q e F_ {\mu \nu}+ g\tilde F _ {\mu \nu})/q =F _ {\mu \nu}\cos \theta +\tilde F _ {\mu \nu} \sin \theta;\nonumber \\
  \tilde  {\cal F}_ {\mu \nu}= (Q e\tilde F _ {\mu \nu}- gF _ {\mu \nu})/q =-F _ {\mu \nu} \sin \theta +\tilde F _ {\mu \nu} \cos \theta,
  \eea
  where $q=\sqrt {Q^2e^2+g^2}$, we come to the Maxwell equations  with the one kind of sources:
  \be \label{max1}
  \partial ^\nu {\cal F}_ {\mu \nu}= q j_\mu ; \quad \partial ^\nu \tilde {\cal F}_ {\mu \nu}=0
  \ee
  and to the ''usual'' Lorentz force acting onto the trial charge $q$,
  \be \label{lof}
  f^\nu =  q j_\mu {\cal F}^{\mu \nu}.
  \ee
  Thus, formally, we have gone over (with the aid of the dual rotations (\ref{dual povorot})) to the usual Maxwell electrodynamics involving a one effective charge $q$ from the electrodynamics involving dually charged particles with the universal ratio $g/Qe$. This means that these both forms are equivalent.
  
  The formal posibility of such going over has a profound physical justification. Since the presence of a field can be discovered only by its impact on a charged body, while the  charge of a particle can be identified, in turn, only with the aid of the field, then only the interaction effects between charges and fields can be treated as  immediately observable (measurable) phenomena, but not {\it charges and fields taken separately}. Therefore, it is impossible, in principle, to establish a difference between the effects (\ref{tt})- (\ref{lf}) and (\ref{max1})- (\ref{lof}) if one identifyes the effective charge $q$, (\ref{dual povorot}), with the obsrvable electric charge.
  
 \medskip But, as it was stressed in \cite{tom}, the dual rotation (\ref{dual povorot}) leads to the electrodynamics with only a one charge for all the sources involving, only when the  ratio $g/Qe$ is the same for all the particles. Otherwise, going over to the system with an effective charge is possible only for the particles of the one kind: say, $q_1=\sqrt{(Q_1e_1)^2+g_1^2}$. The   particles with another ratio $g/Qe$ ($g_2/(Q_2e_2)\neq g_1/(Q_1e_1)$) will possess both the electric, $e'$, and magnetic, $g'$, charges \cite{Zwanz}
 
 \bea 
 \label{Zw}
 e'=q_2\cos (\theta_2- \theta_1); \nonumber\\
 g'=q_2 \sin (\theta_2- \theta_1),
 \eea
 where $q_2=\sqrt {(Q_2e_2)^2+g_2^2}$, $\theta_i=\arctan (g_i/(Q_ie_i))$ ($i=1,2$).
 
 Thus it becomes evident that the universality of the ratio $g/Qe$ purchases a crucial importance in electrodynamics of dually charged particles: if this ratio is the same for all the particles, the observable  magnetic charge is absent.
 
 \bigskip The situation changes drastically when nontrivial topologies are involved: in particular, when one consideres the abelian $U(1)$ gauge theory. 
 
 This becomes obvious at examining the Dirac quantization condition 
 
 \be \label{dirn} \frac {Q_ng_n}{4\pi} =\frac 1 2 n; \quad n\in {\bf Z}, \ee 
 which is the generalization of the standard Dirac quantization condition (\ref{dq}) \cite{Cheng, Ryder, Dirac} onto the case when the electric charge $Q_n$ and the magnetic one, $g_n$, are involved in the $n^{\rm th}$ topological sector of the  abelian $U(1)$ gauge model (for instance, these charges are relevant to some dyon with the topological charge $n$). 
 
 As a consequence of Eq. (\ref{dirn}),
 
 \be \label{dirn1}
 Q_n= n \frac {2\pi}{g_n}\Leftrightarrow \frac {Q_n}{g_n}=\frac {2\pi n}{g^2_n}
 \ee
 {\it and these relations are different in each topological sector}.  This allows for observable magnetic charges to appear in the abelian $U(1)$ gauge theory due to the above arguments \cite{tom}.
 
 \bigskip Existence of nontrivial topologies is the 
distinctive feature of the world in which we live.  In particular, such nontrivial topologies are inherent in the abelian $U(1)\simeq S^1$ gauge theory since $\pi_1 S^1={\bf Z}$. So, to trow these  nontrivial topologies (involving magnetic monopole configurations) and to leave only the  trivial one, $n=0$, which describe correctly Maxwell/quantum electrodynamics, it would be irrational.

  \section{Discussion.}
  In this last section of our study we should like stop on the topic of Section 3, on the author's opinion, the most important in this article \footnote{As to constraint-shell QED, we study in Section 2, it was rather the  
review of that made in the earler papers \cite{Pervush2,Nguen1,Nguen2002}.  The only point deserves here the especial our attention. It is the possibility to represent the Dirac variables ${\bf A}^D(x)$ in the shape (\ref{Direq}) of retarding potentials. This bridges constraint-shell QED \cite{Pervush2,Nguen1,Nguen2002} with the Feynman theory (more specifically, with the $<f\vert {\bf S}^{(2)} \vert i>$  matrix elements involving two fermionic currents)}. 

\medskip In Section 3 the "complete" $U(1)$ gauge theory involving Dirac monopole configurations (\ref{adec})- (\ref{acil1}) \cite{Cheng,Ryder} was discussed and its Dirac fundamental quantization was performed, leading us to the "reduced" Hamiltonian  (\ref{Hamil}).

The principal conclusion which can be drawn from this is that  constraint-shell QED \cite{Pervush2,Nguen1,Nguen2002} involving Dirac variables is, in fact, only  some topologically trivial part of this "complete" theory. In this some analogy can be observed with the "complete" liquid ${\rm He}^4$ model. There, as it is well known, superfluidity in ${\rm He}^4$ \cite{Kapitza,Landau} corresponds to the trivial $n=0$ topology while at $n\neq 0$ vortices arise in a liquid ${\rm He}^4$
specimen (see the monograph \cite{Halatnikov} and the discussion in the recent paper \cite{disc}).

More in detail, the appearance of rectilinear vortices in a liquid helium II specimen is set by Eq.
$$ n= \frac{m}{2\pi \hbar} \oint \limits _\Gamma {\bf v}^{(n)}d {\bf l}; \quad n \in {\bf Z}. $$
This Eq. implicates the helium mass $m$ and the tangential velocity ${\bf v}^{(n)}$ of a rectilinear vortex. At $n=0$, as it can be seen transparently from this Eq.,  the integral in its r.h.s. disappears. This means that the trivial topology $n=0$ corresponds to irrotational, superfluid motions, ${\rm rot}~ {\bf v}^{(n)}=0$, inside the liquid helium II specimen.

\medskip Further, two kinds of (gauge) field configurations were us found with respect to the electric/magnetic charges ascribed to these configurations. For nontrivial topologies $n\neq 0$ there are {\it dyon} configurations, involving electric as well as magnetic charges (such configurations are,  
actually, well known in theoretical physics), while the trivial topology $n= 0$  permites  purely magnetic states (we have called them "`magnons" in the present study).

 \medskip The intersting point of the model us discussed is the possibility of the topological expansion for the Hamiltonian  (\ref{Hamil}). In other words, {\it any energy integral can be  expanded by subintegrals equipped by topological numbers $n$}.
 
 .
 
 
 \bigskip The next intersting example permitting a ``topological'' expansion is the Bogomol'nyi bound \cite{Al.S.}
 $$E_{\rm min}= 4\pi {\bf m} \frac {a}{g},\quad a\equiv \frac{m}{\sqrt{\lambda}}; $$
 for the BPS monopole vacuum configuration (with $m$ and $\lambda$ being the Higgs mass and selfinteraction constant, respectively, taken in the BPS limit  \cite{BPS} $m\to \infty$, $\lambda\to \infty$).
 
 In the latter equation the dependence of $E_{\rm min}$ on the topological number $n$   
originates from the  dependence on $n$ of the magnetic charge $\bf m$. Following \cite{Al.S.}, the latter one can be given (indeed, upon some fitting) as
$$ {\bf m} (\Phi,A)= C~ \zeta (\Phi,A), \quad \zeta (\Phi,A)\in {\bf Z}$$
for a magnetic monopole vacuum (Higgs-YM) configuration $\Phi,A$.

Here $C=\nu/4\pi$, where $\nu$ can be found  from the conditions
$$ \exp (\nu h)=1; \quad \exp (\lambda h)\neq 1 $$
($h\equiv h(\Phi) \equiv \Phi /{a}$) as $0\leq \lambda\leq \nu$. From the geometrical point of view, $\nu$  is characterized as the length of the circle $U(1)\simeq S^1$ (of the unit radius).\par

\medskip If $H$ is the survived symmetry group in the considered model (in the present study we consider $H=U(1)$) and if $t(h)$ is a representation of its Lee algebra, then the operator $t(h)$ has the system of eigenvalues $\{\lambda_k\}$ (since the operator $h$ is anti-Hermitian, its eigenvalues $\{\lambda_k\}$ are imaginary).

It  follows from the relation $\exp (\nu h)=1$ that 
 $$ T(\exp (\nu h))= \exp (\nu~ T(h))=1.$$
Therefore for all the eigenvalues $\lambda_k$ of the operator $t(h)$ the equality $\exp (\nu h)=1$ is satisfied. 

Whence $$ \nu \lambda_k = 2\pi n i, \quad n\in {\bf Z};$$ 

On the other hand, if $\Phi=\sum \limits_k  \Phi^k f_k$, where $f_1,\dots f_n$ are { \it eigenvectors} of the operator $t(h)$ and if only the "electromagnetic" part of the gauge field $A_\mu$: $a_\mu$, is different from zero (i.e. $ A_\mu=a_\mu h$), then 
$$D _\mu \Phi \equiv \partial_\mu \Phi + t(a_\mu h) \Phi =\sum \limits_k (\partial_\mu \Phi^k + \lambda_k a_\mu \Phi^k )f_k.$$
Thus the electric charge entering this covariant derivative is inferred to be $2\pi n/\nu$.

Since magnetic charges are integer multiples of the number $\nu (4\pi)^{-1}$ \cite{Al.S.},  the product of the electric charge $e$ of a particle onto the magnetic charge $\bf m$ of (another) particle is a half-integer:
$$e{\bf m}= \frac{1}{2} n, \quad n \in {\bf Z},$$
in agreement with the Dirac hypothesis \cite{Dirac}.

\bigskip The author recognizes that his study of the Dirac "fundamental" quantization of the Abelian $U(1)$ gauge model is only the first little step in this direction and that a rather large job awaits here.
 
\section{Acknowledges} I am very grateful to Drs. Andrzej Borowiec,  
   Lukasz Andrzej Glinka, Victor Perushin, Mikhail Plyushchay, Christian Rakotonirina for their interest to my recent publications.

\end{document}